\documentclass[aps, pra, onecolumn, superscriptaddress, groupedaddress]{revtex4}

\usepackage[english,italian]{babel}
\usepackage{graphicx}  % needed for figures
\usepackage{dcolumn}   % needed for some tables
\usepackage{bm}        % for math
\usepackage{bbm}        % for math
\usepackage{amssymb}   % for math
\usepackage{rotating}
\usepackage{amsmath}
\usepackage{mathtools}
\usepackage[utf8]{inputenc}
\usepackage[colorlinks,bookmarks=false,citecolor=blue,linkcolor=red,urlcolor=blue]{hyperref}

\usepackage{cleveref}
\crefrangeformat{equation}{#3(#1 #4--#5 #2)#6}

\usepackage{soul} % for striked out text \st{...}
\newcommand{\ignore}[1]{}
\usepackage{enumitem}

\newcommand{\eg}{\textit{e.g. }}				% e.g.
\newcommand{\ie}{\textit{i.e. }}				% i.e.

\usepackage[squaren]{SIunits}
\newcommand{\he}{^3\text{He}}

\newcommand{\avg}[1]{\left\langle #1 \right\rangle}
\newcommand{\ket}[1]{\left\vert #1 \right\rangle}
\newcommand{\bra}[1]{\left\langle #1 \right\vert}

\newcommand{\rgs}{\rho_{\text{f}}}
\newcommand{\rms}{\rho_{\text{m}}}
\newcommand{\Tr}{\text{Tr}}
\newcommand{\Tre}{\text{Tr}_{\text{e}}}
\newcommand{\Trn}{\text{Tr}_{\text{n}}}

\newcommand{\gaf}{\gamma_f}
\newcommand{\gam}{\gamma_m}

\newcommand{\gyrOne}{\gamma_{1/2}}
\newcommand{\gyrThree}{\gamma_{3/2}}
\newcommand{\gyrNuc}{\gamma_{nuc}}

\newcommand{\reToo}{\mathfrak{R}\TT{1}{1}}
\newcommand{\imToo}{\mathfrak{I}\TT{1}{1}}
\newcommand{\reTto}{\mathfrak{R}\TT{2}{1}}
\newcommand{\imTto}{\mathfrak{I}\TT{2}{1}}
\newcommand{\reTtt}{\mathfrak{R}\TT{2}{2}}
\newcommand{\imTtt}{\mathfrak{I}\TT{2}{2}}
\newcommand{\reTro}{\mathfrak{R}\TT{3}{1}}
\newcommand{\imTro}{\mathfrak{I}\TT{3}{1}}
\newcommand{\reTrt}{\mathfrak{R}\TT{3}{2}}
\newcommand{\imTrt}{\mathfrak{I}\TT{3}{2}}
\newcommand{\reTrr}{\mathfrak{R}\TT{3}{3}}
\newcommand{\imTrr}{\mathfrak{I}\TT{3}{3}}
\newcommand{\TT}[2]{T^{#1}_{#2}}

\begin{document}
\selectlanguage{english}

\title{Effective Faraday interaction between light and nuclear spins of Helium-3 in its ground state: a semiclassical study \vspace{10mm}}

\author{Matteo Fadel}
\email[]{fadelm@phys.ethz.ch} 
\affiliation{Department of Physics ETH Z\"urich - 8093 Z\"urich - Switzerland} 
\affiliation{Department of Physics University of Basel - Klingelbergstrasse 82 4056 Basel - Switzerland} 
\author{Philipp Treutlein}
\email[]{philipp.treutlein@unibas.ch} 
\affiliation{Department of Physics University of Basel - Klingelbergstrasse 82 4056 Basel - Switzerland} 
\author{Alice Sinatra}
\email[]{alice.sinatra@lkb.ens.fr}
\affiliation{Laboratoire Kastler Brossel ENS-Universit\'e PSL CNRS Universit\'e de la Sorbonne et Coll\`ege de France - 24 rue Lhomond 75231 Paris - France}

%\date{\today}

\begin{abstract}\vspace{20mm}
We derive the semiclassical evolution equations for a system consisting of helium-3 atoms in the $2^3S$ metastable state interacting with a light field far-detuned from the $2^3S-2^3P$ transition, in the presence of metastability exchange collisions with ground state helium atoms and a static magnetic field. For two configurations, each corresponding to a particular choice of atom-light detuning in which the contribution of either the metastable level $F=1/2$ or $F=3/2$ is dominant, we derive a simple model of three coupled collective spins from which we can analytically extract an effective coupling constant between the collective nuclear spin and light. In these two configurations, we compare the predictions of our simplified model with the full model.
\end{abstract}

\maketitle

\clearpage
\newpage
\tableofcontents

\newpage

\section{Introduction}
 Helium-3 in its ground state has a purely nuclear spin 1/2. Protected by a complete electronic shell and separated from the first excited state by 20 eV, the nuclear spin of helium is a two-level quantum system that offers exceptionally long coherence times of hundreds of hours \cite{Gentile}. The possibility of effectively interfacing it with light, which transports information over long distances and can be measured at the quantum noise level, offers interesting application prospects for quantum technologies \cite{Dantan,Reinaudi,FirstenbergPolzik,KatzQM,AlanLong,AlanPRL}. The Faraday interaction between the collective spin of an ensemble of atoms and the Stokes spin of light provides a light-matter quantum interface that has already been demonstrated in the laboratory in the case of alkaline atoms \cite{KuzmichEPL,HammererRMP10,StroboNat15}. In the case of the purely nuclear spin of rare gases in their ground state, interaction with light requires an intermediate system. For helium-3, it is a small fraction of atoms brought into a metastable state that offers near-infrared transitions and interacts with atoms in the ground state via metastability exchange collisions \cite{Gentile}. In this paper we study in detail the interaction of light with a set of helium-3 atoms, a fraction of which is brought into the metastable state.  
 At the semiclassical level, we explore the validity of the simplified model used in \cite{AlanLong,AlanPRL}, taking into account the full atomic structure in the metastable $3^{3}S$ and the excited $2^{3}P$ states. Moreover, we propose another possible configuration that should allow for a larger effective coupling between the nuclear spins and the light.

\section{Light-matter interaction for metastable helium atoms}

\begin{figure}[h!]
	\centering
		\includegraphics[width=0.6\textwidth]{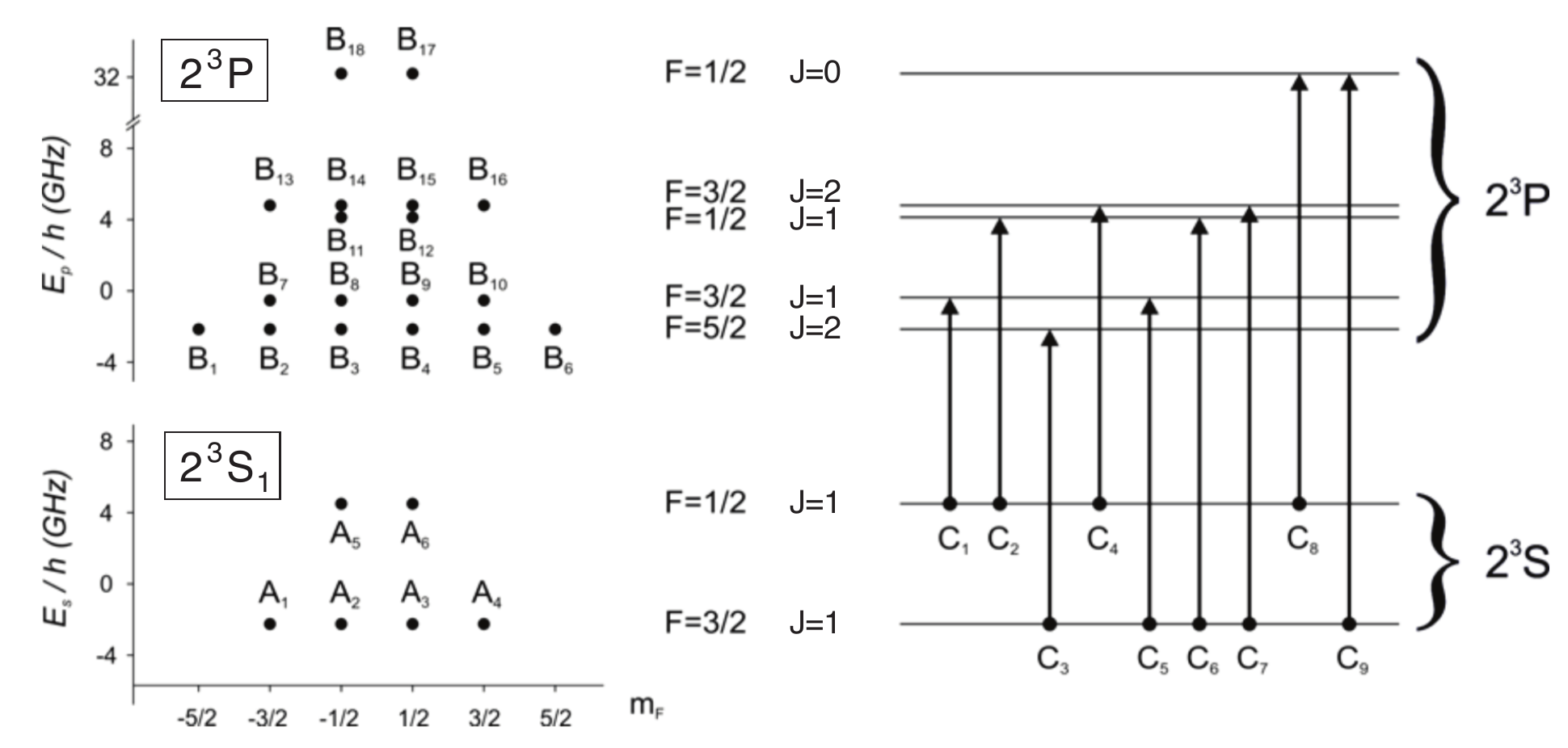}
	\caption{ Left: hyperfine structure of states $2^3$S$_1$ and $2^3$P in $^3\text{He}$. Right: allowed transitions at $\unit{1083}{nm}$.} 	
\label{fig:levels}
\end{figure}

While the ground state $1^1S_0$ of helium-3 is purely nuclear, the metastable state $2^3 S_1$ has an electronic component and is the starting level for transitions at $\unit{1083}{nm}$ to the excited states $2^3 P$. Figure \ref{fig:levels}(Left) shows the hyperfine structure of the metastable and excited levels. The accessible transitions between levels are shown in Figure \ref{fig:levels}(Right), and the relative frequencies in Table \ref{tab:Cfreqs} of Appendix \ref{app:table_fr}.

For light propagating along the $z$-axis, we introduce the components of the Stokes spin in terms of the creation and annihilation operators of a photon polarised in the $x$ or $y$ direction, and in term of the circularly polarized photons creation and annihilation operators $a_1=(a_x-i a_y)/\sqrt{2}$, $a_2=(a_x+i a_y)/\sqrt{2}$
\begin{align}
    S_0 &= ( a_x^\dagger a_x + a_y^\dagger a_y ) / 2 = ( a_1^\dagger a_2 + a_2^\dagger a_1 ) / 2 \\
    S_x &= ( a_x^\dagger a_x - a_y^\dagger a_y ) / 2 = ( a_1^\dagger a_1 + a_2^\dagger a_2 ) / 2\\
    S_y &= ( a_x^\dagger a_y + a_y^\dagger a_x ) / 2 = ( a_1^\dagger a_2 - a_2^\dagger a_1 ) / 2i \\
    S_z &= ( a_x^\dagger a_y - a_y^\dagger a_x ) / 2i  = ( a_2^\dagger a_2 - a_1^\dagger a_1 ) / 2
    \label{eq:Stokes}
\end{align}
In the weak saturation regime, the interaction of light with atoms in either one of the two states $F=1/2$ or $F=3/2$ of metastable helium can be described by an effective Hamiltonian obtained by adiabatically eliminating the optical coherences and the populations of the excited state \cite{Pinard07,KuzmichPRA99,footnoteStokes}. The general form of the effective Hamiltonian for one atom of spin $F$ with light a is recalled in Appendix \ref{app:Heff}. 

\subsection{Effective atom-light interaction in the metastable state}
\begin{figure}[htb]
    \centering
    \includegraphics[width=0.70\textwidth]{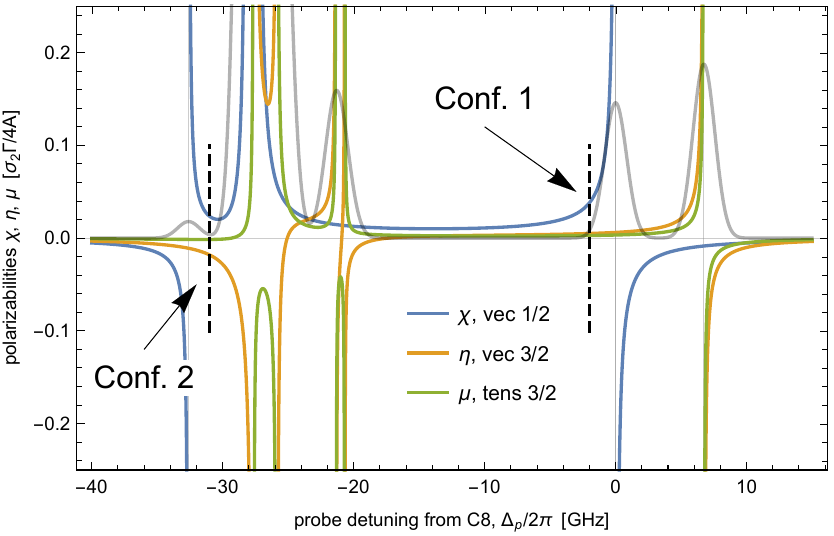}\\
    \bigskip
    \includegraphics[width=0.35\textwidth]{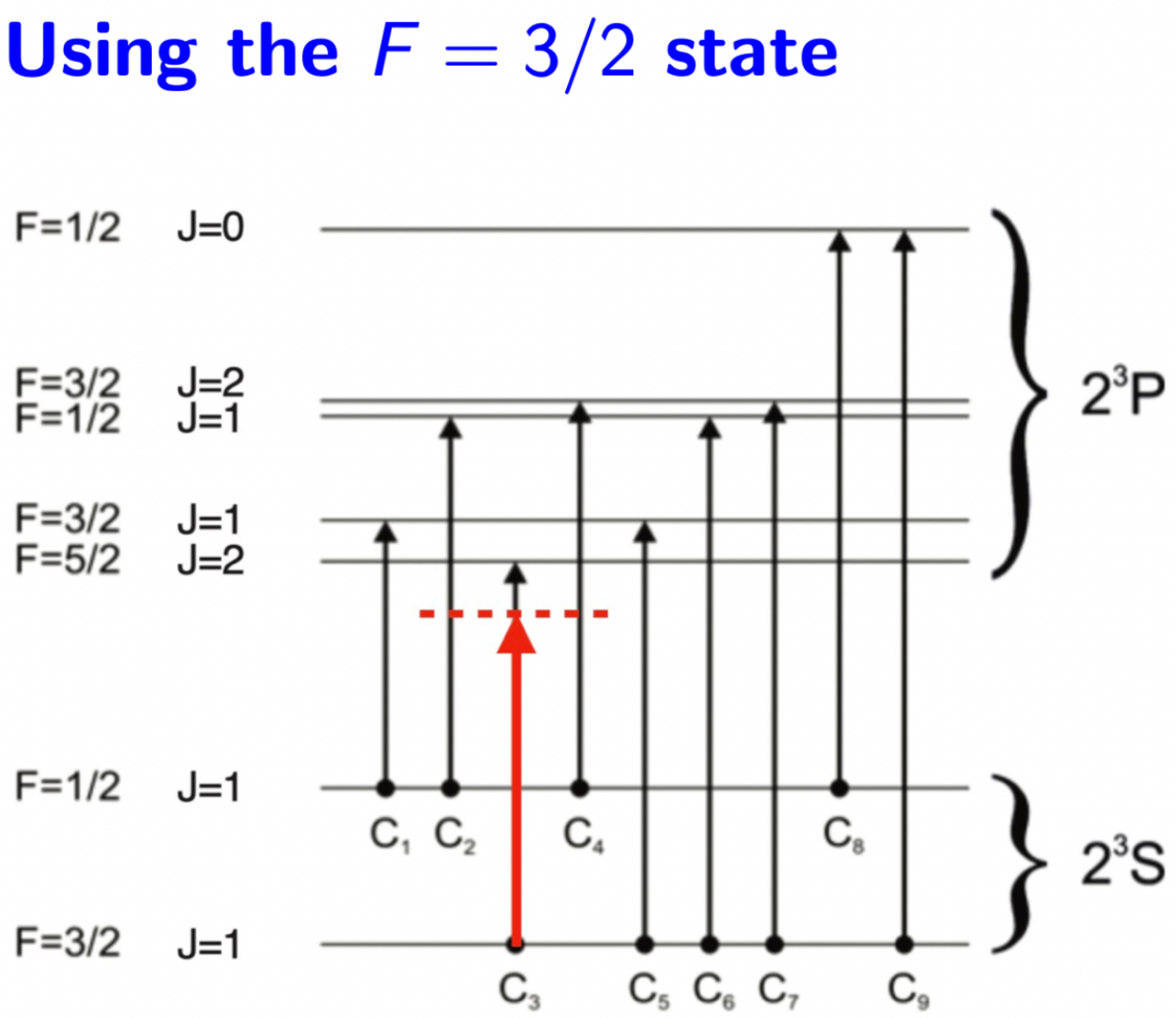}\qquad
    \includegraphics[width=0.35\textwidth]{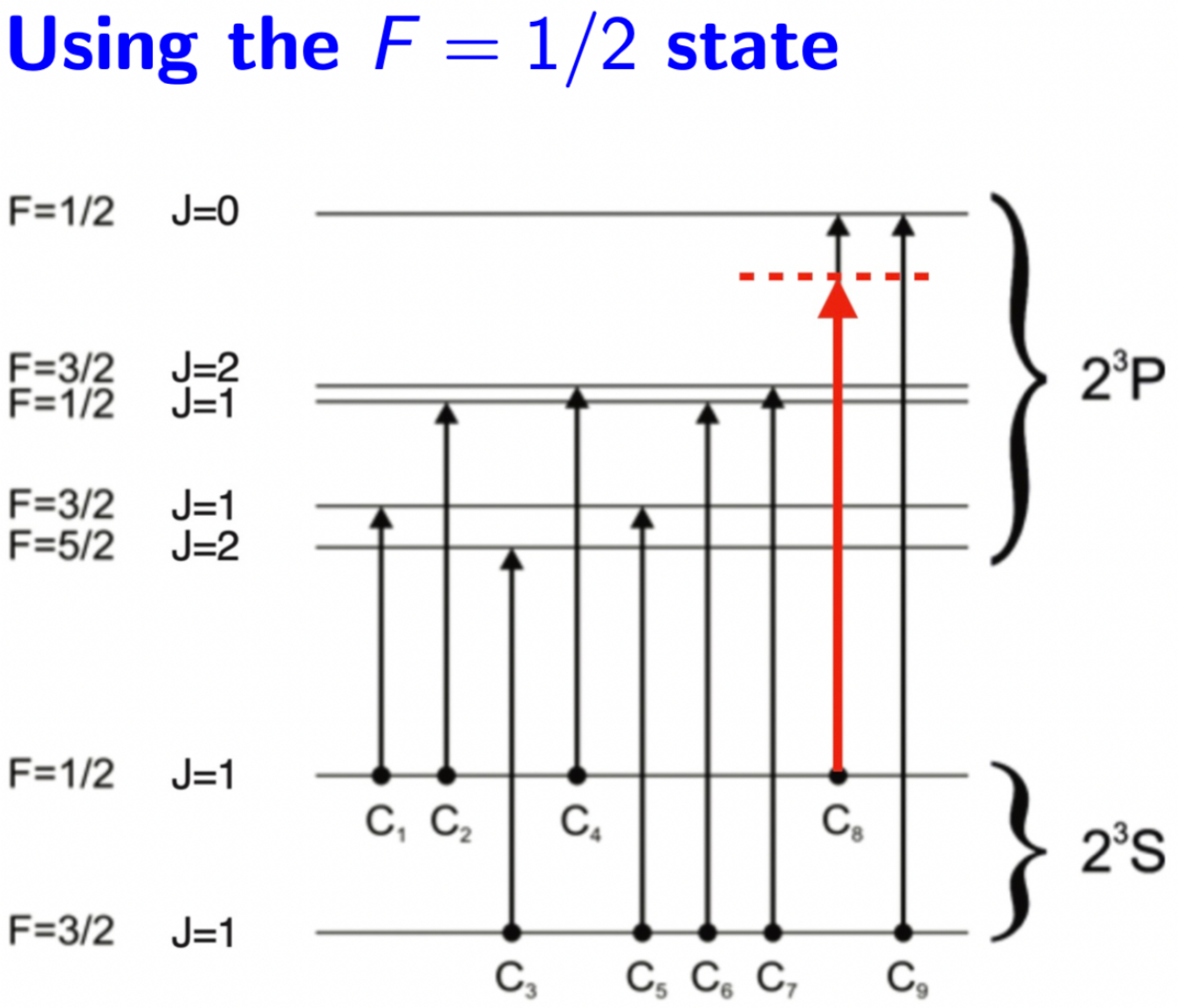}%\\       
    \caption{Top: Coupling constants $\chi$ Eq.~(\ref{eq:chi}) (blue line), $\eta$ Eq.~(\ref{eq:eta}) (orange line) and $\mu$ Eq.~(\ref{eq:mu}) (green line) for the $F=1/2$ and $F=3/2$ levels of the $\he$ metatable, as a function of the light frequency detuning $\Delta_p$ taking the $C_8$ transition as the origin. All the constants are divided by the constant $4A/(\sigma_2\Gamma)$. The grey line shows the absorption spectrum taking into account the Doppler broadening for $T=\unit{300}{K}$ for a non polarized sample \cite{BatzPhD}. Two possible operating point marked as ``Config.2" at $\Delta_p/(2\pi) =\unit{-31}{GHz}$ and ``Config.1" at $\Delta_p/(2\pi)=\unit{-2}{GHz}$ are discussed in the text. Bottom: Level scheme depicting the two possible operating point marked as ``Config.2" and ``Config.1", respectively.} 
    \label{fig:polariz}
\end{figure}
In the case of the metastable state $2^3 S$ of helium-3, we first introduce the collective operators $\vec{K}$, $\vec{J}$ and $\TT{l}{m}$, respectively obtained by summing the single atom spin operators in the $F=1/2$, and spin and tensor operators in the $F=3/2$ manifolds of the metastable state. Considering the transitions to the excited states $2^3 P$, in the case of large detuning, the effective light-atoms hamiltonian for the ensemble then takes the form
\begin{equation}
H_{LA}=H_{1/2}+H_{3/2}^V + H_{3/2}^T \;.
\label{eq:HintL}
\end{equation}
where the two vectorial contributions, of the $F=1/2$ metastable state ($\vec{K}$) and of the $F=3/2$ metastable state ($\vec{J}$) take the Faraday form
\begin{equation}
H_{1/2}=\hbar \chi K_z S_z \; \quad H_{3/2}^V = \hbar \eta J_z S_z \;,
\end{equation}
and the tensorial contribution of the $F=3/2$ metastable state takes the form
\begin{align}
    H_{3/2}^T &= \sum_{i=1}^{n} \hbar \mu \left[ \left( \dfrac{F_i(F_i+1)}{3} - F_{i,z}^2 \right) S_0 + (F_{i,x}^2-F_{i,y}^2) S_x + (F_{i,x} F_{i,y} + F_{i,y} F_{i,x}) S_y \right] \notag \\ 
    &= \hbar\mu \left[ - 2 \TT{2}{0} S_0 + \sqrt{12} \left( \reTtt S_x + \imTtt S_y \right) \right] \;.
\end{align} 
In the above, we used the collective irreducible tensor operators $\TT{2}{0}$, $\reTtt$ and $\imTtt$, that are obtained by summing the single-atom tensor operators defined in Appendix \ref{app:Heff}.

The constants $\chi$, $\eta$ and $\mu$ representing the strength of the different contributions have the form
\begin{eqnarray}
\chi &=& \dfrac{\sigma_2}{4A} \Gamma \left(  \dfrac{2}{9 (\Delta_p - \Delta_{1})} - \dfrac{8}{9 (\Delta_p - \Delta_{2})} + \dfrac{10}{9 (\Delta_p - \Delta_{4})} - \dfrac{4}{9 \Delta_p}   \right) \label{eq:chi} \\
\eta &=& \dfrac{\sigma_2}{4A} \Gamma \left(  \dfrac{3}{5 (\Delta_p - \Delta_{3})} - \dfrac{2}{9 (\Delta_p - \Delta_{5})} - \dfrac{1}{9 (\Delta_p - \Delta_{6})} - \dfrac{2}{45 (\Delta_p - \Delta_{7})} - \dfrac{2}{9 (\Delta_p - \Delta_{9})} \right)  \label{eq:eta} \\
\mu &=& \dfrac{\sigma_2}{10 A} \Gamma \left(  - \dfrac{1}{4 (\Delta_p - \Delta_{3})} + \dfrac{5}{9 (\Delta_p - \Delta_{5})} - \dfrac{5}{36 (\Delta_p - \Delta_{6})} + \dfrac{1}{9 (\Delta_p - \Delta_{7})} - \dfrac{5}{18 (\Delta_p - \Delta_{9})} \right)  \label{eq:mu} \,.
\end{eqnarray} 
In these equations, $\sigma_2=3\lambda^2/2\pi$, $A$ is the cross sectional area of the light mode,  $\Gamma\approx\unit{10^7}{s^{-1}}$ is the excited-state spontaneous decay rate, and taking the $C_8$ transition as a reference, we have defined $\Delta_p=\omega_{\text{probe}}-\omega_{C_8}$ and $\Delta_{i} = \omega_{F F'}-\omega_{C_8}$.
In Figure \ref{fig:polariz} we represent the three coupling constants $\chi$ (\ref{eq:chi}), $\eta$ (\ref{eq:eta}) and $\mu$ (\ref{eq:mu}), all divided by the constant $4A/(\sigma_2\Gamma)$, as a function of light frequency. For $\Delta_p/(2\pi)=\unit{-2}{GHZ}$, which is the operating point considered in \cite{AlanLong,AlanPRL} and marked as ``Config.1" in Fig.~\ref{fig:polariz}, the vector contribution of $F=1/2$ is dominant. A second interesting operating point, marked as ``Config.2" in Fig.~\ref{fig:polariz}, is for $\Delta_p/(2\pi)=\unit{-31}{GHZ}$, around the local minimum of the grey absorption curve where the tensor part of $F=3/2$ is relatively small. The main advantage of this configuration is that one could work with a highly polarized state, $M\simeq1$, for which the $F=1/2$ spin manifold is empty and the initial state is effectively a spin coherent state \cite{spinT}.
\hfill
\newpage

\subsection{Metastable atomic variables evolution due to the interaction with the light}
\label{sub:eomL}
Due to the Hamilonian (\ref{eq:HintL}), we find from $dO/dt=i[H,O]/\hbar$ that the Stokes operators of the light and the collective atomic variable obey the following equation of motion 
{\scriptsize
 \begin{align}
    \dfrac{\text{d} S_x}{\text{d} t} \bigg\vert_{\text{L}} &= - \chi K_z S_y - \eta J_z S_y + \sqrt{12} \mu \imTtt S_z \\
    \dfrac{\text{d} S_y}{\text{d} t} \bigg\vert_{\text{L}} &= \chi K_z S_x + \eta J_z S_x - \sqrt{12} \mu \reTtt S_z \\
    \dfrac{\text{d} S_z}{\text{d} t} \bigg\vert_{\text{L}} &= \sqrt{12} \mu \left( \reTtt S_y - \imTtt S_x \right) \label{eq:LMSzEOM}\\
    \dfrac{\text{d} K_x}{\text{d} t} \bigg\vert_{\text{L}} &=  - \chi K_y S_z \\
    \dfrac{\text{d} K_y}{\text{d} t} \bigg\vert_{\text{L}} &= \chi K_x S_z \\
    \dfrac{\text{d} K_z}{\text{d} t} \bigg\vert_{\text{L}} &= 0 \\
    \dfrac{\text{d} J_x}{\text{d} t} \bigg\vert_{\text{L}} &= -\eta J_y S_z + \sqrt{12} \mu \left( \imTto (S_x - S_0) - \reTto S_y \right) \\
    \dfrac{\text{d} J_y}{\text{d} t} \bigg\vert_{\text{L}} &= \eta J_x S_z + \sqrt{12} \mu \left( \reTto (S_x + S_0) + \imTto S_y \right) \\
    \dfrac{\text{d} J_z}{\text{d} t} \bigg\vert_{\text{L}} &= 2\sqrt{12} \mu \left( \imTtt S_x - \reTtt S_y \right) \\
    \dfrac{\text{d}}{\text{d} t} \reTtt \bigg\vert_{\text{L}} &= - 2 \eta S_z \imTtt + \sqrt{12} \mu \left( \dfrac{1}{\sqrt{5}} S_y (2\TT{1}{0}-\TT{3}{0}) + \dfrac{1}{\sqrt{3}} S_0 \imTrt \right) \\
    \dfrac{\text{d}}{\text{d} t} \imTtt \bigg\vert_{\text{L}} &= 2 \eta S_z \reTtt - \sqrt{12} \mu \left( \dfrac{1}{\sqrt{5}} S_x (2\TT{1}{0}-\TT{3}{0}) + \dfrac{1}{\sqrt{3}} S_0 \reTrt \right) \\ 
    \dfrac{\text{d}}{\text{d} t} \reTto \bigg\vert_{\text{L}} &= - \eta S_z \imTto + \sqrt{6}\mu \left(  S_x\left(-\sqrt{\dfrac{3}{5}}\imTro - \dfrac{\sqrt{2}}{5}J_y + \imTrr \right) 
    %+ \right. \\ & \left. 
    + S_0\left( \dfrac{2}{\sqrt{15}}\imTto - \dfrac{\sqrt{2}}{5} J_y \right) + S_y \left( \sqrt{\dfrac{3}{5}}\reTro + \dfrac{\sqrt{2}}{5}J_x - \reTrr \right) \right) \\
    \dfrac{\text{d}}{\text{d} t} \imTto \bigg\vert_{\text{L}} &= \eta S_z \reTto + \sqrt{6}\mu \left(  S_x\left(\sqrt{\dfrac{3}{5}}\reTro + \dfrac{\sqrt{2}}{5}J_x + \reTrr \right) 
    %+ \right. \\ & \left. 
    + S_0\left( \dfrac{2}{\sqrt{15}}\reTto - \dfrac{\sqrt{2}}{5} J_x \right) + S_y \left( \sqrt{\dfrac{3}{5}}\imTro + \dfrac{\sqrt{2}}{5}J_y + \imTrr \right) \right) \\
    \dfrac{\text{d}}{\text{d} t} \TT{2}{0} \bigg\vert_{\text{L}} &= \sqrt{12}\mu \left( S_x \imTrt - S_y \reTrt \right) \\
    \dfrac{\text{d}}{\text{d} t} \reTrr \bigg\vert_{\text{L}} &= - 3 \eta S_z \imTrr + \sqrt{6} \mu \left( S_x \imTto + S_y \reTto \right) \\
    \dfrac{\text{d}}{\text{d} t} \imTrr \bigg\vert_{\text{L}} &= 3 \eta S_z \reTrr - \sqrt{6} \mu \left( S_x \reTto - S_y \imTto \right) \\
    \dfrac{\text{d}}{\text{d} t} \reTrt \bigg\vert_{\text{L}} &= - 2 \eta S_z \imTrt + 2 \mu \left( \sqrt{3} S_y \TT{2}{0} + S_0 \imTtt \right) \\
    \dfrac{\text{d}}{\text{d} t} \imTrt \bigg\vert_{\text{L}} &= 2 \eta S_z \reTrt - 2 \mu \left( \sqrt{3} S_x \TT{2}{0} + S_0 \reTtt \right) \\
    \dfrac{\text{d}}{\text{d} t} \reTro \bigg\vert_{\text{L}} &= - \eta S_z \imTro + \sqrt{\dfrac{2}{5}} \mu \left( (2 S_0 + 3 S_x) \imTto - 3 S_y \reTto \right) \\
    \dfrac{\text{d}}{\text{d} t} \imTro \bigg\vert_{\text{L}} &= \eta S_z \reTro - \sqrt{\dfrac{2}{5}} \mu \left( (2 S_0 - 3 S_x) \reTto - 3 S_y \imTto \right) \\
    \dfrac{\text{d}}{\text{d} t} \TT{3}{0} \bigg\vert_{\text{L}} &= -\dfrac{2}{5}\sqrt{15} \mu \left( S_x \imTtt - S_y \reTtt \right) 
\end{align}
}
\section{External magnetic field}
\label{sub:eomB}
 In the presence of a static magnetic field $\vec{B}$, the system evolves according to the Hamiltonian \begin{align}
    H_B &= - \hbar \left( \gyrOne \vec{B}\cdot\vec{K} + \gyrThree \vec{B}\cdot\vec{J} + \gyrNuc \vec{B}\cdot\vec{I} \right) \,, \qquad \mbox{where} \\
    \gyrOne &=  \dfrac{4}{3} \gamma_{ms} \; ;\quad
    \gyrThree = \dfrac{2}{3} \gamma_{ms} \; ;\quad
    \gamma_{ms} = - 2\pi\, \unit{2.802}{MHz/G} \; ;\quad
    \gyrNuc = - 2\pi\, \unit{3.243}{kHz/G} \; ;
    \label{eq:gyro}
\end{align} 
are the gyromagnetic ratios \cite{DRpaper1}.

\subsection{Metastable atomic variables evolution due to an external magnetic field}
The corresponding equations of motion for atomic variables are given by 
{\scriptsize
 \begin{align}
    \dfrac{\text{d} \vec{I}}{\text{d} t}\bigg\vert_{B} &= \gyrNuc \; \vec{I} \times \vec{B}  \\
    \dfrac{\text{d} \vec{K}}{\text{d} t}\bigg\vert_{B} &= \gyrOne \; \vec{K} \times \vec{B} \\
    \dfrac{\text{d} \vec{J}}{\text{d} t}\bigg\vert_{B} &= \gyrThree \; \vec{J} \times \vec{B} \\
    \dfrac{\text{d}}{\text{d} t} \reTtt \bigg\vert_{B} &= \gyrThree \left( B_x \imTto + B_y \reTto + 2 B_z \imTtt \right) \\
    \dfrac{\text{d}}{\text{d} t} \imTtt \bigg\vert_{B} &= \gyrThree \left( -B_x \reTto + B_y \imTto - 2 B_z \reTtt \right) \\
    \dfrac{\text{d}}{\text{d} t} \reTto \bigg\vert_{B} &= \gyrThree \left( B_x \imTtt + B_y \left(\sqrt{3}\TT{2}{0} - \reTtt \right) + B_z \imTto \right) \\
    \dfrac{\text{d}}{\text{d} t} \imTto \bigg\vert_{B} &= \gyrThree \left( -B_x \left(\sqrt{3}\TT{2}{0} + \reTtt \right) - B_y \imTtt - B_z \reTto \right) \\ 
    \dfrac{\text{d}}{\text{d} t} \TT{2}{0} \bigg\vert_{B} &= \gyrThree \sqrt{3} \left( B_x\imTto - B_y \reTto \right) \\
    \dfrac{\text{d}}{\text{d} t} \reTrr \bigg\vert_{B} &= \gyrThree \left( B_x \sqrt{\dfrac{3}{2}}\imTrt + B_y \sqrt{\dfrac{3}{2}}\reTrt + B_z 3 \imTrr \right) \\
    \dfrac{\text{d}}{\text{d} t} \imTrr \bigg\vert_{B} &= \gyrThree \left( -B_x \sqrt{\dfrac{3}{2}}\reTrt + B_y \sqrt{\dfrac{3}{2}}\imTrt - B_z 3 \reTrr \right) \\
    \dfrac{\text{d}}{\text{d} t} \reTrt \bigg\vert_{B} &= \gyrThree \left( B_x \left( \sqrt{\dfrac{10}{4}}\imTro + \sqrt{\dfrac{3}{2}}\imTrr \right) + B_y \left( \sqrt{\dfrac{10}{4}}\reTro - \sqrt{\dfrac{3}{2}}\reTrr \right) + B_z 2 \imTrt \right) \\
    \dfrac{\text{d}}{\text{d} t} \imTrt \bigg\vert_{B} &= \gyrThree \left( B_x \left(- \sqrt{\dfrac{10}{4}}\reTro - \sqrt{\dfrac{3}{2}}\reTrr \right) + B_y \left( \sqrt{\dfrac{10}{4}}\imTro - \sqrt{\dfrac{3}{2}}\imTrr \right) - B_z 2 \reTrt \right) \\
        \dfrac{\text{d}}{\text{d} t} \reTro \bigg\vert_{B} &= \gyrThree \left( B_x \sqrt{\dfrac{5}{2}} \imTrt + B_y \left(\sqrt{6} \TT{3}{0} - \sqrt{\dfrac{5}{2}} \reTrt \right) +B_z \imTro \right) \\
        \dfrac{\text{d}}{\text{d} t} \imTro \bigg\vert_{B} &= \gyrThree \left( - B_x \left(\sqrt{6} \TT{3}{0} - \sqrt{\dfrac{5}{2}} \reTrt \right) - B_y \sqrt{\dfrac{5}{2}} \imTrt - B_z \reTro \right) \\
        \dfrac{\text{d}}{\text{d} t} \TT{3}{0} \bigg\vert_{B} &= \gyrThree \sqrt{6} \left( B_x  \imTto - B_y \reTto \right) 
\end{align}
}

\section{Metastability exchange collisions}

\subsection{Evolution of the one-body density matrix}
Metastability exchange collisions (MEC) couple the metastable state to the ground state of helium. They are usually described in terms of the one-body density matrix $\rho$ which is assumed to be block-diagonal, with the $2\times 2$ matrix $\rgs$ describing the ground state and the $6\times 6$ matrix $\rms$ describing the metastable state. Following a collision, $\rho$ transforms according to $\rho \xrightarrow{\text{MEC}} \rho^\prime$ with \cite{DRpaper1,Reinaudi} \begin{align}
    \rgs^\prime &= \Tre [\rms] \label{eq:rhoPT1}\\
    \rms^\prime &= \rgs \otimes \Trn [\rms] \label{eq:rhoPT2}
\end{align} where $\Tre$ and $\Trn$ denote the trace on the electronic and nuclear degrees of freedom respectively. Note that the electronic degrees of freedom do not appear in $\rgs$ since the ground state is a singlet state, \ie $S=0$.

Considering a set of $N_{\rm cell}$ atoms in the ground state and $n_{\rm cell}$ in the metastable state, the equations of motion of the one-body density matrix are written \begin{align}
    \dfrac{d}{dt} \rgs &= \dfrac{1}{T} \left( - \rgs + \Tre [\rms] \right) \label{eq:rhoMEC1}\\
    \dfrac{d}{dt} \rms &= \dfrac{1}{\tau} \left( - \rms + \rgs \otimes \Trn [\rms] \right) \label{eq:rhoMEC2}
\end{align}
where the two collision rates $\gaf=1/T$ et $\gam=1/\tau$ for an atom in the ground and metastable states, respectively, satisfy the relation \begin{equation}
   \dfrac{\gam}{\gaf} = \dfrac{T}{\tau} = \dfrac{N_{\rm cell}}{n_{\rm cell}}  \;.
   \label{eq:gam_gaf}
\end{equation} 
Such equations of motion, expressed in the $\{\ket{i}\}$ basis of the Zeeman sublevels of helium-3, are found in~\cite{Reinaudi}. They allow us to calculate the evolution due to the exchange of any one-body atomic operator $O$ \begin{equation}\label{eq:eomOpMEC}
    \dfrac{d \avg{O}}{dt}\bigg \vert_{\text{MEC}} = \Tr\left[ O \dfrac{d \rho}{dt} \bigg \vert_{\text{MEC}} \right] \;.
\end{equation} 
A detailed example explaining how we proceed to obtain the equations in given in Appendix \ref{app:deivMEC}.

\subsection{Atomic variables evolution due to metastability exchange}
\label{sub:eomMEC}

For the three spins: ground $I$, metastable $(F=1/2)$ $K$ and metastable $(F=3/2)$ $J$, we obtain the semi-classical equations \begin{align}
    \dfrac{\text{d} \avg{\vec{I}}}{\text{d} t}\bigg\vert_{\text{MEC}} &= - \dfrac{1}{T} \avg{\vec{I}} + \dfrac{1}{3T} \dfrac{N}{n} \left( \avg{\vec{J}} - \avg{\vec{K}} \right) \\
    \dfrac{\text{d} \avg{\vec{K}}}{\text{d} t}\bigg\vert_{\text{MEC}} &= - \dfrac{7}{9 \tau} \avg{\vec{K}} + \dfrac{1}{9\tau} \avg{\vec{J}} - \dfrac{1}{9\tau} \dfrac{n}{N} \avg{\vec{I}} - \dfrac{4}{3\tau} \dfrac{1}{N} \avg{\vec{\vec{Q}}} \cdot \avg{\vec{I}} \\
    \dfrac{\text{d} \avg{\vec{J}}}{\text{d} t}\bigg\vert_{\text{MEC}} &= - \dfrac{4}{9 \tau} \avg{\vec{J}} + \dfrac{10}{9\tau} \avg{\vec{K}} + \dfrac{10}{9\tau} \dfrac{n}{N} \avg{\vec{I}} + \dfrac{4}{3\tau} \dfrac{1}{N} \avg{\vec{\vec{Q}}} \cdot \avg{\vec{I}} \;,
\end{align} 
where we have introduced the collective alignment tensor $\avg{Q_{\alpha\beta}} = \sum_{i=1}^N \frac{1}{3} \frac{1}{6}\left( \frac{3}{2} \avg{F_{i,\alpha} F_{i,\beta} + F_{i,\beta} F_{i,\alpha}} - 2 \delta_{\alpha\beta} \right) $ \cite{DRpaper1}

%where we have introduced for convenience the alignment tensor $\avg{Q_{ij}} = \frac{1}{6}\left( \frac{3}{2} \avg{\Sigma_i \Sigma_j + \Sigma_j \Sigma_i} - 2 \delta_{ij} \right) $ and the electron spin operator expectation value in the metastable $\avg{\vec{\Sigma}} = \frac{2}{3}\left( \avg{\vec{J}} + 2 \avg{\vec{K}} \right)$.  Using the Wigner-Eckart theorem, we can write $\avg{Q_{ij}} = \frac{1}{3} \frac{1}{6}\left( \frac{3}{2} \avg{J_i J_j + J_j J_i} - 2 \delta_{ij} \right) $ \cite{DRpaper1} and identify 
\begin{subequations}
\begin{align}
    Q_{xx} &= \dfrac{1}{6} \left( \sqrt{3} \reTtt - \TT{2}{0} \right) \\
    Q_{yy} &= -\dfrac{1}{6} \left( \sqrt{3} \reTtt + \TT{2}{0} \right) \\
    Q_{zz} &= \dfrac{1}{3} \TT{2}{0} \\
    Q_{xy} &= \dfrac{1}{2\sqrt{3}} \imTtt \\
    Q_{xz} &= -\dfrac{1}{2\sqrt{3}} \reTto \\
    Q_{yz} &= -\dfrac{1}{2\sqrt{3}} \imTto \;.
\end{align}
\end{subequations} 

The rank-2 tensors of state $F=3/2$ evolve according to equations \begin{align}
    \dfrac{\text{d}}{\text{d} t} \avg{\reTtt}\bigg\vert_{\text{MEC}} &= - \dfrac{2}{3 \tau} \avg{\reTtt} + \dfrac{1}{\sqrt{3}\tau} \dfrac{1}{N} \left( \avg{I_x}\avg{\Sigma_x} - \avg{I_y}\avg{\Sigma_y} \right) \\
    \dfrac{\text{d}}{\text{d} t} \avg{\imTtt}\bigg\vert_{\text{MEC}} &= - \dfrac{2}{3 \tau} \avg{\imTtt} + \dfrac{1}{\sqrt{3}\tau} \dfrac{1}{N} \left( \avg{I_x}\avg{\Sigma_y} + \avg{I_y}\avg{\Sigma_x} \right) \\
    \dfrac{\text{d}}{\text{d} t} \avg{\reTto}\bigg\vert_{\text{MEC}} &= - \dfrac{2}{3 \tau} \avg{\reTto} + \dfrac{1}{\sqrt{3}\tau} \dfrac{1}{N} \left( \avg{I_x}\avg{\Sigma_z} + \avg{I_z}\avg{\Sigma_x} \right) \\
    \dfrac{\text{d}}{\text{d} t} \avg{\imTto}\bigg\vert_{\text{MEC}} &= - \dfrac{2}{3 \tau} \avg{\imTto} + \dfrac{1}{\sqrt{3}\tau} \dfrac{1}{N} \left( \avg{I_y}\avg{\Sigma_z} + \avg{I_z}\avg{\Sigma_y} \right) \\
    \dfrac{\text{d} \avg{\TT{2}{0}}}{\text{d} t}\bigg\vert_{\text{MEC}} &= - \dfrac{2}{3 \tau} \avg{\TT{2}{0}} + \dfrac{1}{3\tau} \dfrac{1}{N} \left( 3 \avg{I_z}\avg{\Sigma_z} - \avg{\vec{I}} \cdot \avg{\vec{\Sigma}} \right)
\end{align}
where we defined the electron spin operator expectation value in the metastable state $\avg{\vec{\Sigma}} = \frac{2}{3}\left( \avg{\vec{J}} + 2 \avg{\vec{K}} \right)$.

The rank-3 tensors of state $F=3/2$ evolve according to equations 
\begin{align}
    \dfrac{\text{d}}{\text{d} t} \avg{\reTrr}\bigg\vert_{\text{MEC}} &= - \dfrac{1}{\tau} \avg{\reTrr} - \dfrac{\sqrt{6}}{3 \tau}\dfrac{1}{N} \left( \avg{I_x}\avg{\reTtt} - \avg{I_y}\avg{\imTtt} \right) \\
    \dfrac{\text{d}}{\text{d} t} \avg{\imTrr}\bigg\vert_{\text{MEC}} &= - \dfrac{1}{\tau} \avg{\imTrr} - \dfrac{\sqrt{6}}{3 \tau}\dfrac{1}{N} \left( \avg{I_x}\avg{\imTtt} - \avg{I_y}\avg{\reTtt} \right) \\
    \dfrac{\text{d}}{\text{d} t} \avg{\reTrt}\bigg\vert_{\text{MEC}} &= - \dfrac{1}{\tau} \avg{\reTrt} - \dfrac{2}{3 \tau}\dfrac{1}{N} \left( \avg{I_x}\avg{\reTto} - \avg{I_y}\avg{\imTto} - \avg{I_z}\avg{\reTtt} \right) \\
    \dfrac{\text{d}}{\text{d} t} \avg{\imTrt}\bigg\vert_{\text{MEC}} &= - \dfrac{1}{\tau} \avg{\imTrt} - \dfrac{2}{3 \tau}\dfrac{1}{N} \left( \avg{I_x}\avg{\imTto} + \avg{I_y}\avg{\reTto} - \avg{I_z}\avg{\imTtt} \right) \\
    \dfrac{\text{d}}{\text{d} t} \avg{\reTro}\bigg\vert_{\text{MEC}} &= - \dfrac{1}{\tau} \avg{\reTro} + \dfrac{2}{3 \sqrt{10}\tau}\dfrac{1}{N} \left( \avg{I_x}\left(\avg{\reTtt}-2\sqrt{3}\avg{\TT{2}{0}} \right) + \avg{I_y}\avg{\imTtt} + 4 \avg{I_z}\avg{\reTto} \right) \\
    \dfrac{\text{d}}{\text{d} t} \avg{\imTro}\bigg\vert_{\text{MEC}} &= - \dfrac{1}{\tau} \avg{\imTro} + \dfrac{2}{3 \sqrt{10}\tau}\dfrac{1}{N} \left( \avg{I_x}\avg{\imTtt} - \avg{I_y} \left(\avg{\reTtt}-2\sqrt{3}\avg{\TT{2}{0}} \right) + 4 \avg{I_z}\avg{\imTto} \right) \\
    \dfrac{\text{d}}{\text{d} t} \avg{\TT{3}{0}}\bigg\vert_{\text{MEC}} &= - \dfrac{1}{\tau} \avg{\TT{3}{0}} + \dfrac{2}{\sqrt{5}\tau}\dfrac{1}{N} \left( \dfrac{2\sqrt{3}}{6} \left( \avg{I_x}\avg{\reTto} + \avg{I_y}\avg{\imTto} \right) + \avg{I_z}\avg{\TT{2}{0}} \right) 
\end{align} 

\section{Semiclassical equations of motion}
\subsection{Semiclassical equations for the atomic and field variables}
Using the results of sections \ref{sub:eomL}, \ref{sub:eomB} and \ref{sub:eomMEC}, we can write the semiclassical equations of motion for the averages of the Stokes and atomic collective spin components operators, under the influence of (i) the light-atom interaction in the metastable state, (ii) a uniform external magnetic field, and (iii) metastability exchange collisions. 
The term ``semiclassical" means here that all the operators, including those in equations of sections \ref{sub:eomL} and \ref{sub:eomB}, are replaced by their expectation values. The time derivative of a semiclassical variable $\langle O \rangle$ has three contributions:
\begin{equation}\label{eq:EOMfull}
    \dfrac{\text{d} \langle O \rangle}{\text{d} t} = \dfrac{\text{d} \langle O \rangle}{\text{d} t} \bigg\vert_{\text{L}} + \dfrac{\text{d} \langle O \rangle}{\text{d} t} \bigg\vert_{\text{B}}  + \dfrac{\text{d} \langle O \rangle}{\text{d} t} \bigg\vert_{\text{MEC}} \;.
\end{equation}

\subsection{Stationary solution}
\label{sec:stat}
For a nuclear polarisation $M\in [-1,1]$, a fixed magnetic field along the $x$-direction, $\vec{B}=B_x \vec{e}_x$, and a fixed light intensity and polarisation, the semi-classical equations of motion have a stationary solution that is found by setting the time derivatives to zero. Here we consider the Stokes spin and the nuclear spin polarised along the static magnetic field in the $x$-direction,
 \begin{equation}
 \label{eq:stat0}
    \avg{S_x}_s = \dfrac{n_{ph}}{2} \;,\qquad \avg{S_y}_s = \avg{S_z}_s = 0 \;,\qquad \avg{I_x}_s = M\dfrac{N}{2}\;,\qquad \avg{I_y}_s = \avg{I_z}_s = 0 \;,
\end{equation} The stationary solution for the atomic variables in the metastable state and its small polarization expansion is then: \begin{subequations}\label{eq:stat1}
\begin{align}
    \avg{K_x}_s &= \dfrac{M}{2} \left(\dfrac{1-M^2}{3+M^2}\right) n_{\rm cell} \stackrel{M\to0}{\simeq} \left( \dfrac{M}{6} - \dfrac{2 M^3}{9} + \mathcal{O}(M^5) \right) n_{\rm cell}\\
    \avg{J_x}_s &= M \left(\dfrac{5+M^2}{3+M^2}\right) n_{\rm cell} \stackrel{M\to0}{\simeq}\left( \dfrac{5 M}{3} - \dfrac{2 M^3}{9} + \mathcal{O}(M^5) \right) n_{\rm cell} \\
    \avg{\TT{2}{0}}_s &= - \left(\dfrac{M^2}{3+M^2}\right) n_{\rm cell} \stackrel{M\to0}{\simeq}\left( -\dfrac{M^2}{3} + \mathcal{O}(M^4) \right) n_{\rm cell} \\
    \avg{\reTtt}_s &= \sqrt{3} \left(\dfrac{M^2}{3+M^2}\right) n_{\rm cell} \stackrel{M\to0}{\simeq} \left( \dfrac{M^2}{\sqrt{3}} + \mathcal{O}(M^4) \right) n_{\rm cell}\\
    \avg{\reTro}_s &= \sqrt{\dfrac{3}{10}} \left(\dfrac{M^3}{3+M^2}\right) n_{\rm cell} \stackrel{M\to0}{\simeq}\left( \dfrac{M^3}{\sqrt{30}} + \mathcal{O}(M^5) \right) n_{\rm cell} \\
    \avg{\reTrr}_s &= -\dfrac{1}{\sqrt{2}} \left(\dfrac{M^3}{3+M^2}\right) n_{\rm cell} \stackrel{M\to0}{\simeq}\left( -\dfrac{M^3}{3\sqrt{2}} + \mathcal{O}(M^5) \right) n_{\rm cell} \;.
\end{align}
\end{subequations} 
For $M\to0$, the quantities (\ref{eq:stat1}a,b), (\ref{eq:stat1}c,d) and (\ref{eq:stat1}e,f) are respectively of order one, two and three in $M$, indicating that the tensor contributions can be neglected for a small polarisation.

Moreover, we note that the stationary solutions Eqs.~(\ref{eq:stat1}a-d) are identical to those found in Ref.~\cite{AlanLong}, although the light-matter interaction for the $F=3/2$ level of the metastable state was not included in that work.

\subsection{Linearised equations of motion}
The linearised semiclassical equations around the stationary solution (\ref{eq:stat1}) are obtained by substituting $\langle O \rangle \rightarrow \avg{O}_s + \delta O$, where $\delta O$ is the small variation of $\langle O \rangle$ from its steady state value, and keeping only linear terms in the variations. Introducing the fluctuation vector $\vec{c}=(\vec{a}, \vec{b})$ with
\begin{eqnarray}
    \vec{a} &=& (\delta S_y, \delta S_z, \delta I_y, \delta I_z, \delta K_y, \delta K_z, \delta J_y, \delta J_z, \delta \reTto, \delta \imTtt, \delta \TT{3}{0}, \delta \imTro, \delta \reTrt, \delta \imTrr ) \\
    \vec{b} &=& (\delta S_0, \delta S_x, \delta I_x, \delta K_x, \delta J_x,   \delta \TT{2}{0}, \delta \imTto, \reTtt, \delta \reTro, \delta \imTrt, \delta \reTrr) \;,
\end{eqnarray}
the linearised equations of motion take the block-diagonal form
\begin{equation} \label{eq:linEOMfull}
      \dot{\vec{c}} = \left[\begin{array}{ c | c }
    A & 0 \\
    \hline
    0 & B
  \end{array}\right] \vec{c} \;,
\end{equation}
where the first block represents fluctuations in the $yz$-plane, namely the plane perpendicular to the spin polarisation.
In the expression of the matrices $A$ and $B$ given below, the lines isolate the sub-blocks for the light, nuclear and metastable states respectively.

\begin{sideways}
\begin{minipage}{\textheight}
{\tiny
\begin{equation}
    A = \left(
\begin{array}{cc|cc|cccccccccc}
 0 & -\frac{6 \mu  M^2 n}{M^2+3} & 0 & 0 & 0 & \frac{\chi  n_{\text{ph}}}{2} & 0 & \frac{\eta  n_{\text{ph}}}{2} & 0 & 0 & 0 & 0 & 0 & 0 \\
 \frac{6 \mu  M^2 n}{M^2+3} & 0 & 0 & 0 & 0 & 0 & 0 & 0 & 0 & -\sqrt{3} \mu  n_{\text{ph}} & 0 & 0 & 0 & 0 \\
 \\
 \hline
 \\
 0 & 0 & -\frac{1}{T} & B_x \gamma _{\text{nuc}} & -\frac{N}{3 n T} & 0 & \frac{N}{3 n T} & 0 & 0 & 0 & 0 & 0 & 0 & 0 \\
 0 & 0 & -B_x \gamma _{\text{nuc}} & -\frac{1}{T} & 0 & -\frac{N}{3 n T} & 0 & \frac{N}{3 n T} & 0 & 0 & 0 & 0 & 0 & 0 \\
 \\
 \hline
 \\
 0 & \frac{1}{2} M \left(\frac{4}{M^2+3}-1\right) n \chi  & -\frac{n-M^2 n}{3 M^2 N \tau +9 N \tau } & 0 & -\frac{7}{9 \tau } & \gamma _{\text{1/2}} B_x & \frac{1}{9 \tau } & 0 & 0 & -\frac{M}{3 \sqrt{3} \tau } & 0 & 0 & 0 & 0 \\
 0 & 0 & 0 & -\frac{n-M^2 n}{3 M^2 N \tau +9 N \tau } & \gamma _{\text{1/2}} \left(-B_x\right) & -\frac{7}{9 \tau } & 0 & \frac{1}{9 \tau } & \frac{M}{3 \sqrt{3} \tau } & 0 & 0 & 0 & 0 & 0 \\
 0 & \frac{\eta  M \left(M^2+5\right) n}{M^2+3} & \frac{2 \left(M^2+5\right) n}{3 \left(M^2+3\right) N \tau } & 0 & \frac{10}{9 \tau } & 0 & -\frac{4}{9 \tau } & \gamma _{\text{3/2}} B_x & 2 \sqrt{3} \mu  n_{\text{ph}} & \frac{M}{3 \sqrt{3} \tau } & 0 & 0 & 0 & 0 \\
 -\frac{12 \mu  M^2 n}{M^2+3} & 0 & 0 & \frac{2 \left(M^2+5\right) n}{3 \left(M^2+3\right) N \tau } & 0 & \frac{10}{9 \tau } & \gamma _{\text{3/2}} \left(-B_x\right) & -\frac{4}{9 \tau } & -\frac{M}{3 \sqrt{3} \tau } & 2 \sqrt{3} \mu  n_{\text{ph}} & 0 & 0 & 0 & 0 \\
 \frac{2 \sqrt{3} \mu  M \left(M^2+1\right) n}{M^2+3} & 0 & 0 & -\frac{4 M n}{\sqrt{3} \left(M^2 N \tau +3 N \tau \right)} & 0 & -\frac{2 M}{3 \sqrt{3} \tau } & -\frac{2}{5} \sqrt{3} \mu  n_{\text{ph}} & -\frac{M}{3 \sqrt{3} \tau } & -\frac{2}{3 \tau } & \gamma
   _{\text{3/2}} B_x & 0 & -\frac{\mu  n_{\text{ph}}}{\sqrt{10}} & 0 & \sqrt{\frac{3}{2}} \mu  n_{\text{ph}} \\
 0 & \frac{2 \sqrt{3} \eta  M^2 n}{M^2+3} & \frac{4 M n}{\sqrt{3} \left(M^2 N \tau +3 N \tau \right)} & 0 & \frac{2 M}{3 \sqrt{3} \tau } & 0 & \frac{M}{3 \sqrt{3} \tau } & -\frac{2}{5} \sqrt{3} \mu  n_{\text{ph}} & \gamma _{\text{3/2}} \left(-B_x\right) & -\frac{2}{3
   \tau } & \sqrt{\frac{3}{5}} \mu  n_{\text{ph}} & 0 & -\mu  n_{\text{ph}} & 0 \\
 \frac{6 \mu  M^2 n}{\sqrt{5} \left(M^2+3\right)} & 0 & 0 & -\frac{2 M^2 n}{\sqrt{5} \left(M^2 N \tau +3 N \tau \right)} & 0 & 0 & 0 & 0 & \frac{M}{\sqrt{15} \tau } & -\sqrt{\frac{3}{5}} \mu  n_{\text{ph}} & -\frac{1}{\tau } & \sqrt{6} \gamma _{\text{3/2}} B_x & 0 & 0
   \\
 0 & \frac{\sqrt{\frac{3}{10}} \eta  M^3 n}{M^2+3} & \frac{\sqrt{\frac{2}{15}} M^2 n}{M^2 N \tau +3 N \tau } & 0 & 0 & 0 & 0 & 0 & \frac{\mu  n_{\text{ph}}}{\sqrt{10}} & \frac{M}{3 \sqrt{10} \tau } & -\sqrt{6} \gamma _{\text{3/2}} B_x & -\frac{1}{\tau } &
   -\sqrt{\frac{5}{2}} \gamma _{\text{3/2}} B_x & 0 \\
 -\frac{2 \sqrt{3} \mu  M^2 n}{M^2+3} & 0 & 0 & \frac{2 M^2 n}{\sqrt{3} \left(M^2 N \tau +3 N \tau \right)} & 0 & 0 & 0 & 0 & -\frac{M}{3 \tau } & \mu  n_{\text{ph}} & 0 & \sqrt{\frac{5}{2}} \gamma _{\text{3/2}} B_x & -\frac{1}{\tau } & \sqrt{\frac{3}{2}} \gamma
   _{\text{3/2}} B_x \\
 0 & -\frac{3 \eta  M^3 n}{\sqrt{2} \left(M^2+3\right)} & -\frac{\sqrt{2} M^2 n}{M^2 N \tau +3 N \tau } & 0 & 0 & 0 & 0 & 0 & -\sqrt{\frac{3}{2}} \mu  n_{\text{ph}} & -\frac{M}{\sqrt{6} \tau } & 0 & 0 & -\sqrt{\frac{3}{2}} \gamma _{\text{3/2}} B_x & -\frac{1}{\tau } \\
\end{array}
\right)
\end{equation}}

{\tiny
\begin{equation}
    B= \left(
\begin{array}{cc|cc|ccccccc}
 0 & 0 & 0 & 0 & 0 & 0 & 0 & 0 & 0 & 0 & 0 \\
 0 & 0 & 0 & 0 & 0 & 0 & 0 & 0 & 0 & 0 & 0 \\
 \\
 \hline
 \\
 0 & 0 & -\frac{1}{T} & -\frac{N}{3 n T} & \frac{N}{3 n T} & 0 & 0 & 0 & 0 & 0 & 0 \\
 0 & 0 & -\frac{3 M^2 n+n}{3 M^2 N \tau +9 N \tau } & -\frac{7}{9 \tau } & \frac{1}{9 \tau } & \frac{M}{9 \tau } & 0 & -\frac{M}{3 \sqrt{3} \tau } & 0 & 0 & 0 \\
 \\
 \hline
 \\
 0 & 0 & \frac{6 M^2 n+10 n}{3 M^2 N \tau +9 N \tau } & \frac{10}{9 \tau } & -\frac{4}{9 \tau } & -\frac{M}{9 \tau } & 0 & \frac{M}{3 \sqrt{3} \tau } & 0 & 0 & 0 \\
 \\
 0 & 0 & -\frac{4 M n}{3 M^2 N \tau +9 N \tau } & -\frac{2 M}{9 \tau } & -\frac{M}{9 \tau } & -\frac{2}{3 \tau } & \sqrt{3} \gamma _{\text{3/2}} B_x & 0 & 0 & \sqrt{3} \mu  n_{\text{ph}} & 0 \\
 \frac{2 \sqrt{3} \mu  M n}{M^2+3} & -\frac{2 \sqrt{3} \mu  M n}{M^2+3} & 0 & 0 & 0 & -\sqrt{3} \gamma _{\text{3/2}} B_x & -\frac{2}{3 \tau } & \gamma _{\text{3/2}} \left(-B_x\right) & -\sqrt{\frac{5}{2}} \mu  n_{\text{ph}} & 0 & -\sqrt{\frac{3}{2}} \mu  n_{\text{ph}} \\
 0 & 0 & \frac{4 M n}{\sqrt{3} \left(M^2 N \tau +3 N \tau \right)} & \frac{2 M}{3 \sqrt{3} \tau } & \frac{M}{3 \sqrt{3} \tau } & 0 & \gamma _{\text{3/2}} B_x & -\frac{2}{3 \tau } & 0 & \mu  n_{\text{ph}} & 0 \\
 \\
 0 & 0 & \frac{\sqrt{\frac{6}{5}} M^2 n}{M^2 N \tau +3 N \tau } & 0 & 0 & -\frac{\sqrt{\frac{2}{15}} M}{\tau } & \sqrt{\frac{5}{2}} \mu  n_{\text{ph}} & \frac{M}{3 \sqrt{10} \tau } & -\frac{1}{\tau } & \sqrt{\frac{5}{2}} \gamma _{\text{3/2}} B_x & 0 \\
 \\
 -\frac{2 \sqrt{3} \mu  M^2 n}{M^2+3} & \frac{2 \sqrt{3} \mu  M^2 n}{M^2+3} & 0 & 0 & 0 & -\sqrt{3} \mu  n_{\text{ph}} & -\frac{M}{3 \tau } & -\mu  n_{\text{ph}} & -\sqrt{\frac{5}{2}} \gamma _{\text{3/2}} B_x & -\frac{1}{\tau } & -\sqrt{\frac{3}{2}} \gamma _{\text{3/2}}
   B_x \\
 0 & 0 & -\frac{\sqrt{2} M^2 n}{M^2 N \tau +3 N \tau } & 0 & 0 & 0 & \sqrt{\frac{3}{2}} \mu  n_{\rm{ph}} & -\frac{M}{\sqrt{6} \tau } & 0 & \sqrt{\frac{3}{2}} \gamma _{\text{3/2}} B_x & -\frac{1}{\tau } \\
\end{array}
\right)
\end{equation}
}

\end{minipage}
\end{sideways}

\section{Effective coupling between the nuclear spin and the light in two configurations}
\label{sec:configs}

A simplified model involving only three coupled spins (one for the light field, one for the metastable state and one for the fundamental state) can be derived when focusing on one of the two choices of the light frequency detuning shown in figure \ref{fig:polariz}: ``Config.1'' or ``Config.2'', where the vector Hamiltonian of the metastable level $F=1/2$ or the metastable level $F=3/2$ dominates. For this purpose, for a given light detuning, we set the non-dominant interaction terms in $H_{LA}$ Eq.~(\ref{eq:HintL}) to zero and, among the degrees of freedom of the metastable state, we adiabatically eliminate those that evolve only under the influence of the magnetic field and the metastable exchange collisions.  

\subsection{Configuration 1: Exploiting the interaction with the $F=1/2$ manifold}

We start from the linearized equations \eqref{eq:linEOMfull}, and neglect the coupling of the light with the $F=3/2$ manifold by setting $\eta=\mu=0$. Then, we adiabatically eliminate the $\delta J_\alpha$ and $\delta Q_{\alpha x}$ degrees of freedom by solving the algebraic equations $d (\delta J_\alpha)/ dt =0$ and $d (\delta Q_{\alpha x}) / dt =0$, and inserting the solution in the equations of motion for the remaining variables. In terms of the complex variables
\begin{equation}
    I_+ = I_y + i I_z \;,\qquad\qquad  K_+ = K_y + i K_z  \;,
\end{equation}
we obtain
\begin{subequations}\label{eq:conf1AdiabEl}
\begin{align}
    \dfrac{d}{dt} \delta S_y &= \avg{S_x} \chi \delta K_z \\
    \dfrac{d}{dt} \delta S_z &= 0 \\
    \dfrac{d}{dt} \delta I_+ &= - \gamma_f^{(1/2)} \left(\dfrac{a_1^{(1/2)}}{c^{(1/2)}} + i \dfrac{B_x \gamma_{nuc}}{\gamma_f^{(1/2)}} \right) \delta I_+ + \gamma_m^{(1/2)} \dfrac{a_2^{(1/2)}}{c^{(1/2)}} \delta K_+ \\
    \dfrac{d}{dt} \delta K_+ &= - \gamma_m^{(1/2)} \left(\dfrac{b_1^{(1/2)}}{c^{(1/2)}} + i \dfrac{B_x \gamma_{1/2}}{\gamma_f^{(1/2)}} \right) \delta K_+ + \gamma_f^{(1/2)} \dfrac{b_2^{(1/2)}}{c^{(1/2)}} \delta I_+ + \avg{K_x} \chi \delta S_z
\end{align}
\end{subequations}
where we introduced the rescaled polarization-dependent metastability exchange rates
\begin{equation}
    \gamma_f^{(1/2)} = \dfrac{1}{T} \dfrac{(4+M^2)(1-M^2)}{(8-M^2)(3+M^2)} \;,\qquad
    \gamma_m^{(1/2)} = \dfrac{1}{\tau} \dfrac{(4+M^2)}{(8-M^2)} \;,
\end{equation}
and the dimensionless coefficients $a_i^{(1/2)}$, $b_i^{(1/2)}$ and $c^{(1/2)}$, that can be found in Appendix \ref{app:coeffsAdiabEl}. To first order in the product $B_x \gamma_{\text{ms}}/\gamma_m$, where $\gamma_{\text{ms}}$ Eq.~(\ref{eq:gyro}) is the gyromagnetic factor in the metastable state and $\gamma_m$ (\ref{eq:gam_gaf}) is the metastability exchange rate for a metastable atom, one has
\begin{align}
    \dfrac{a_1^{(1/2)}}{c^{(1/2)}} &\stackrel{B_x\to0}{\simeq} 1 - i \frac{6 \left(M^4+37 M^2+60\right)}{
   \left(M^2-8\right)^2 \left(M^2-1\right)} \dfrac{B_x \gamma_{\text{3/2}}}{\gamma_m}\\
   \dfrac{a_2^{(1/2)}}{c^{(1/2)}} &\stackrel{B_x\to0}{\simeq} 1 - i \frac{30  \left(M^2+4\right)}{\left(M^2-8\right)^2} \dfrac{B_x \gamma_{\text{3/2}}}{\gamma_m}\\
   \dfrac{b_1^{(1/2)}}{c^{(1/2)}} &\stackrel{B_x\to0}{\simeq} 1 - i \frac{2 \left(M^4+17 M^2-20\right)}{\left(M^2-8\right)^2} \dfrac{B_x \gamma_{\text{3/2}}}{\gamma_m} \\
   \dfrac{b_2^{(1/2)}}{c^{(1/2)}} &\stackrel{B_x\to0}{\simeq} 1 - i \frac{30 \left(M^2+4\right)}{\left(M^2-8\right)^2} \dfrac{B_x \gamma_{\text{3/2}}}{\gamma_m}\;.
\end{align}

\begin{figure}
    \centering
    \includegraphics[width=\textwidth]{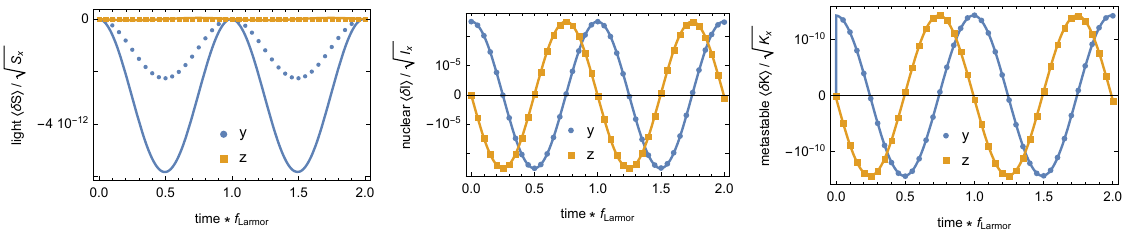}
    \includegraphics[width=\textwidth]{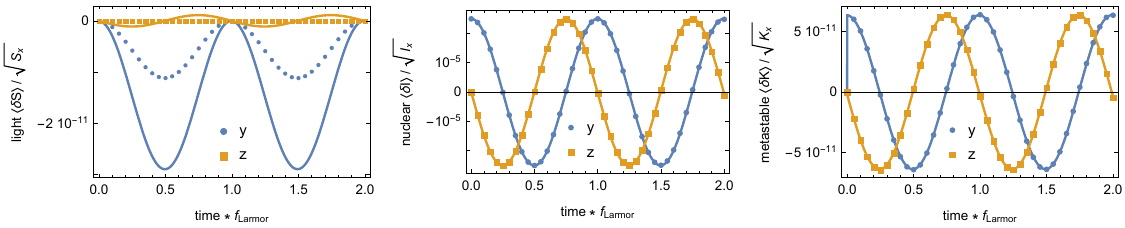}
    \caption{Comparison between the full model Eq.~\eqref{eq:EOMfull} and the simplified model Eq.~\eqref{eq:conf1AdiabEl} in ``Config.1". Top row is for a nuclear magnetization of $M=2\%$, while the bottom row is for $M=10\%$. Blue indicates the $y$ spin component, while yellow the $z$ spin component. Solid lines are the full model, while round and square markers indicate the numerical solution of Eq.~\eqref{eq:conf1AdiabEl}.
    Simulation parameters: $n_\text{ph}/N_{\rm cell}=10^{-3}$, $N_{\text{cell}}/n_{\text{cell}} = T/\tau=\unit{10^{6}}{s}$, and an initial state resulting from tilting the nuclear spin by $\unit{0.01}{rad}$. Time is in units of the nuclear spin Larmor frequency.
    %Simulation parameters: $B_x=\unit{5}{mG}$, $n_\text{ph}=10^3$, $T=\unit{1}{s}$, $\tau=\unit{10^{-6}}{s}$, $N_{\rm cell}=10^6$, and an initial state resulting from tilting the nuclear spin by $\unit{0.01}{rad}$.
    }
    \label{fig:CompConf1}
\end{figure}

We show in Fig.~\eqref{fig:CompConf1} the comparison between the numerical solution of Eqs.~\eqref{eq:conf1AdiabEl} and the full systems of equations Eq.~\eqref{eq:EOMfull} for two values of the nuclear spin polarization, $M=0.02$ and $M=0.1$. While these two models are in good agreement for the atomic variables, which shows the validity of adiabatic elimination there is a discrepancy for the variables $S_y$ and $S_z$ representing the light field. Such a discrepancy is due to the fact that the interaction of light with the $F=3/2$ level, although detuned, is never completely negligible. This is especially visible for the larger polarizations, where the $F=3/2$ manifold is more populated, and for the evolution of the $S_z$ quadrature Eq.~\eqref{eq:LMSzEOM}, which depends exclusively on the tensorial interaction.

In the spirit of Refs.~\cite{AlanLong, AlanPRL}, we are now interested in extracting the effective coupling constant between light and nuclear spin in the limit $B_x \to 0$. This can be done by adiabatically eliminating also the equations for $\delta K$, and then introducing the bosonic quadratures
\begin{subequations}\label{eqPrimakoff}
    \begin{align}
        & \dfrac{\delta S_y}{\sqrt{\avg{S_x}_s}} \simeq X_S \qquad \dfrac{\delta I_y}{\sqrt{\avg{I_x}_s}} \simeq X_I \\
        & \dfrac{\delta S_z}{\sqrt{\avg{S_x}_s}} \simeq P_S \qquad \dfrac{\delta I_z}{\sqrt{\avg{I_x}_s}} \simeq P_I  \;,
    \end{align}
\end{subequations}
satisfying the canonical commutation relations $[X_S,P_S]=[X_I,P_I]=i\hbar$. The resulting equation of motion for the light field is
\begin{equation}\label{eq:eomXS}
    \dfrac{d}{dt} X_S = \chi \dfrac{\avg{K_x}_s}{\avg{I_x}_s} \sqrt{\avg{S_x}_s\avg{I_x}_s} P_S \;,
\end{equation}
from which we obtain the effective Hamiltonian describing an interaction between light and nuclear spin 
\begin{equation}\label{eq:effHPSPI12}
    H_{\text{eff}} = \hbar \Omega^{(1/2)}  P_S P_I \;,
\end{equation}
with the effective coupling rate
\begin{align}
\label{eq:Omega12}
    \Omega^{(1/2)} &= \chi \dfrac{\avg{K_x}_s}{\avg{I_x}_s} \sqrt{\avg{S_x}_s\avg{I_x}_s} \\
%    &= \chi \dfrac{n}{N} \dfrac{\sqrt{n_{\text{ph}} N}}{2} \\
    &= \chi \dfrac{n_{\text{cell}}}{N_{\text{cell}}} \sqrt{n_{\text{ph}} N_{\text{cell}}} f^{(1/2)}(M) \;,
\end{align}
where the second line is obtained by inserting the stationary values Eqs.~(\ref{eq:stat0},\ref{eq:stat1}), and in the last line we defined the polarization-dependent function
\begin{equation}
f^{(1/2)}(M) = \left(\dfrac{1-M^2}{3+M^2}\right)\sqrt{M}  \;.
\label{eq:f12}
\end{equation}

\subsection{Configuration 2: Exploiting the interaction with the $F=3/2$ manifold}\label{sec:Conf2}

For this configuration we want to exploit the interaction of the light with the $F=3/2$ metastable manifold. Therefore, for a large nuclear spin polarization, we neglect the coupling of the light with the $F=1/2$ manifold by setting $\chi=0$ in the linearized equations \eqref{eq:linEOMfull}. In addition, we see from Fig.~\ref{fig:polariz} that in ``Config.2" the tensor polarizability is small, which motivates us to set $\mu=0$ as well. Then we adiabatically eliminate the $\delta K_\alpha$ and $\delta Q_{\alpha x}$ degrees of freedom by solving the algebraic equations $d (\delta K_\alpha)/ dt =0$ and $d (\delta Q_{\alpha x})/ dt =0$ and inserting the solution into the equations of motion for the remaining variables. In terms of the complex variables
\begin{equation}
    I_+ = I_y + i I_z \;,\qquad\qquad  J_+ = J_y + i J_z  \;,
\end{equation}
we obtain
\begin{subequations} \label{eq:conf2AdiabEl}
\begin{align}
    \dfrac{d}{dt} \delta S_y &= \avg{S_x} \eta \delta J_z \\
    \dfrac{d}{dt} \delta S_z &= 0 \\
    \dfrac{d}{dt} \delta I_+ &= - \gamma_f^{(3/2)} \left(\dfrac{a_1^{(3/2)}}{c^{(3/2)}} + i \dfrac{B_x \gamma_{nuc}}{\gamma_f^{(3/2)}} \right) \delta I_+ + \gamma_m^{(3/2)} \dfrac{a_2^{(3/2)}}{c^{(3/2)}} \delta J_+ + \dfrac{a_3^{(3/2)}}{c^{(3/2)}} \avg{J_x} \eta \delta S_z \\
    \dfrac{d}{dt} \delta J_+ &= - \gamma_m^{(3/2)} \left(\dfrac{b_1^{(3/2)}}{c^{(3/2)}} + i \dfrac{B_x \gamma_{3/2}}{\gamma_f^{(3/2)}} \right) \delta J_+ + \gamma_f^{(3/2)} \dfrac{b_2^{(3/2)}}{c^{(3/2)}} \delta I_+ + \left( \dfrac{b_3^{(3/2)}}{c^{(3/2)}} + 1 \right) \avg{J_x} \eta \delta S_z
\end{align}
\end{subequations}
where we introduced the rescaled polarization-dependent metastability exchange rates
\begin{equation}
    \gamma_f^{(3/2)} = \dfrac{1}{T} \dfrac{(4+M^2)(5+M^2)}{(7+M^2)(3+M^2)} \;,\qquad
    \gamma_m^{(3/2)} = \dfrac{1}{\tau} \dfrac{(4+M^2)}{2(7+M^2)} \;.
\end{equation}
and the dimensionless coefficients $a_i^{(3/2)}$, $b_i^{(3/2)}$ and $c^{(3/2)}$, that can be found in Appendix \ref{app:coeffsAdiabEl}. To first order in the product $B_x \gamma_{\text{ms}}\tau$, where $\gamma_{\text{ms}}$ Eq.~(\ref{eq:gyro}) is the gyromagnetic factor in the metastable state and $\tau$ is the inverse of the metastability exchange rate, one has
\begin{align}
    \dfrac{a_1^{(3/2)}}{c^{(3/2)}} &\stackrel{B_x\to0}{\simeq} 1 + i\frac{3 B_x \left(6 \left(M^2+1\right) \gamma _{\text{1/2}}+\left(M^2+13\right) M^2
   \gamma _{\text{3/2}}\right)}{4 \left(M^2+5\right) \left(M^2+7\right)^2 \gamma _m} \\
    \dfrac{a_2^{(3/2)}}{c^{(3/2)}} &\stackrel{B_x\to0}{\simeq} 1 - i \frac{3 B_x \left(2 \left(M^2-2\right) \gamma _{\text{1/2}}+3 M^2 \gamma
   _{\text{3/2}}\right)}{4 \left(M^2+7\right)^2 \gamma _m} \\
    \dfrac{a_3^{(3/2)}}{c^{(3/2)}} &\stackrel{B_x\to0}{\simeq} \frac{3 M^2 }{4 \left(M^2+5\right)
   \left(M^2+7\right)^3} \left(4 \left(M^2+7\right)^2  - 3 i \left(M^2+4\right) \frac{B_x}{\gamma _m} \left(6 \gamma
   _{\text{1/2}}+7 \gamma _{\text{3/2}}\right)\right)\\
    \dfrac{b_1^{(3/2)}}{c^{(3/2)}} &\stackrel{B_x\to0}{\simeq} 1 - i \frac{B_x \left(2 \left(M^4+8 M^2-20\right) \gamma _{\text{1/2}}+9 M^2 \gamma
   _{\text{3/2}}\right)}{4 \left(M^2+7\right)^2 \gamma _m} \\
    \dfrac{b_2^{(3/2)}}{c^{(3/2)}} &\stackrel{B_x\to0}{\simeq} 1 + i \frac{3 B_x \left(2 \left(M^2+1\right) \left(M^2+10\right) \gamma
   _{\text{1/2}}+\left(M^2+13\right) M^2 \gamma _{\text{3/2}}\right)}{4 \left(M^2+5\right)
   \left(M^2+7\right)^2 \gamma _m} \\
    \dfrac{b_3^{(3/2)}}{c^{(3/2)}} &\stackrel{B_x\to0}{\simeq} \frac{3 M^2}{4
   \left(M^2+5\right) \left(M^2+7\right)^3} \left(i 3 \left(M^2+4\right) \frac{B_x}{\gamma _m} \left(2 \left(M^2+10\right) \gamma
   _{\text{1/2}}+7 \gamma _{\text{3/2}}\right)- 4 \left(M^2+7\right)^2 \right)  \;.
\end{align}

\begin{figure}
    \centering
    \includegraphics[width=\textwidth]{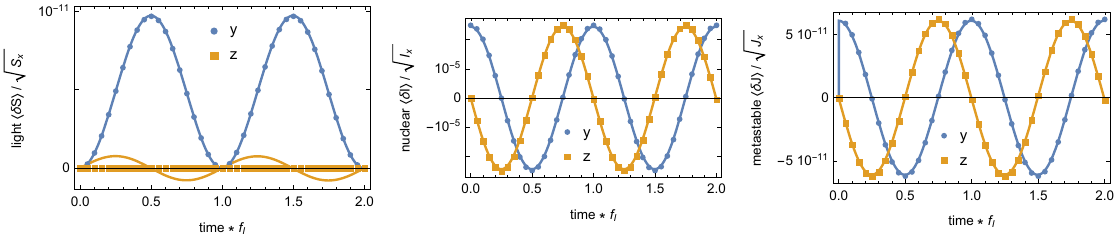}
    \caption{Comparison between the full model Eq.~\eqref{eq:EOMfull} and the simplified model Eq.~\eqref{eq:conf2AdiabEl} in ``Config.2". The nuclear magnetization is set to $M=98\%$. Blue indicates the $y$ spin component, while yellow the $z$ spin component. Solid lines are the full model, while round and square markers indicate the numerical solution of Eq.~\eqref{eq:conf2AdiabEl}.
    Simulation parameters: $n_\text{ph}/N_{\rm cell}=10^{-3}$, $N_{\text{cell}}/n_{\text{cell}} = T/\tau=\unit{10^{6}}{s}$, and an initial state resulting from tilting the nuclear spin by $\unit{0.01}{rad}$. Time is in units of the nuclear spin Larmor frequency.
    %Simulation parameters: $B_x=\unit{5}{mG}$, $n_\text{ph}=10^3$, $T=\unit{1}{s}$, $\tau=\unit{10^{-6}}{s}$, $N=10^6$, $n=1$, and an initial state resulting from tilting the nuclear spin by $\unit{0.01}{rad}$.
    }
    \label{fig:CompConf2}
\end{figure}

We show in Fig.~\eqref{fig:CompConf2} the comparison between the numerical solution of Eqs.~\eqref{eq:conf2AdiabEl} and the full systems of equations Eq.~\eqref{eq:EOMfull} for a nuclear spin polarization of $M=0.98\%$. While these two models are in good agreement for the atomic and $S_y$ variables, there is a discrepancy for the light $S_z$ variable.
As in the previous case, such a discrepancy is due to the residual tensor interaction, as it can be noted by the oscillation of $S_z$ at twice the Larmor frequency and it is more pronounced for large nuclear polarisations due to the $~M^3$ scaling of the tensor components (\ref{eq:stat1}). 
On the other hand, the adiabatic elimination of the $F=1/2$ and tensor degrees of freedom (without setting $\mu$ and $\chi$ to zero) gives a very good approximation of the dynamics, as we show numerically in Appendix \ref{app:adiabNum}.

The effective coupling constant between light and nuclear spin in the $B_x=0$ limit can be extracted similarly to the previous case, now adiabatically eliminating also the equations for $\delta J$. The effective Hamiltonian is then
\begin{equation}\label{eq:effHPSPI32}
    H_{\text{eff}} = \hbar \Omega^{(3/2)}  P_S P_I \;,
\end{equation}
with the effective coupling rate between the light field and the nuclear spin given by
\begin{align}
\label{eq:Omega32}
    \Omega^{(3/2)} &= \eta \dfrac{\avg{J_x}_s}{\avg{I_x}_s} \sqrt{\avg{S_x}_s\avg{I_x}_s} \\
%    &= 3 \eta \dfrac{n}{N} \dfrac{\sqrt{n_{ph} N}}{2} \\
    &= \eta \dfrac{n_{cell}}{N_{cell}} \sqrt{n_{ph} N_{cell}} f^{(3/2)}(M) \;.
\end{align}
Here we used $T/\tau=N_{\text{cell}}/n_{\text{cell}}$, together with the stationary solution for the spins Eqs.~(\ref{eq:stat0},\ref{eq:stat1}), and we defined the polarization-dependent scaling function 
\begin{equation}\label{eq:f32}
    f^{(3/2)}(M)=2 \left(\dfrac{5+M^2}{3+M^2}\right) \sqrt{M} \;.
\end{equation}
In the next section we will compare this result with the one obtained for ``Config.1".

\subsection{Effective coupling in the two configurations and comparison with the full model}

Equations (\ref{eq:Omega12}) and (\ref{eq:Omega32}) for the rates $\Omega^{(1/2)}$ and $\Omega^{(3/2)}$ describing the effective coupling between the collective nuclear spin and the light in ``Config.1" and ``Config.2" respectively, were obtained from approximate models. In this section, we extract such coupling constants from numerical simulations of the full semiclassical equations, and compare the results with the analytical expressions. 

From the evolution of the Stokes spin fluctuation $X_S$, Eqs.~(\ref{eqPrimakoff}-\ref{eq:eomXS}), we see that an oscillation of the collective nuclear spin fluctuation $P_I = P_I(0) \cos(\omega_I t+\phi)$ results in a light signal $X_S=(\Omega P_I(0)/\omega_I) \sin(\omega_I t+\phi)$. Computing the ratio between the oscillation amplitude of the light and nuclear spin gives us $\Omega/\omega_I$, from which we extract the effective coupling for different nuclear polarizations. 

\begin{figure}
    \centering
    \includegraphics[width=0.7\textwidth]{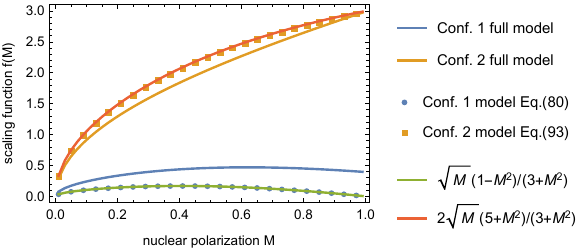}
    \caption{\textbf{Comparison between the effective light-nuclear spin coupling factor in ``Config.1" and ``Config.2" as a function of the nuclear spin magnetization $\mathbf{M}$.} Blue indicates ``Config.1", while yellow ``Config.2". Solid lines refer to the coupling extracted from the full model, while round and square markers refer to the coupling extracted from Eq.~\eqref{eq:conf1AdiabEl} and Eq.~\eqref{eq:conf2AdiabEl}, respectively. Red and green lines are the analytical expressions obtained for $B_x=0$, namely Eq.~\eqref{eq:f12} and Eq.~\eqref{eq:f32} respectively. 
    Simulation parameters: $n_\text{ph}/N_{\rm cell}=10^{-3}$ and $N_{\text{cell}}/n_{\text{cell}} = T/\tau=\unit{10^{6}}{s}$.
    %Simulation parameters: $B_x=\unit{1}{mG}$, $n_\text{ph}=10^3$, $T=\unit{1}{s}$, $\tau=\unit{10^{-6}}{s}$, $N=10^6$, $n=1$, and an initial state resulting from tilting the nuclear spin by $\unit{0.01}{rad}$.
    }
    \label{fig:CompCoup}
\end{figure}

We plot in Figure~\ref{fig:CompCoup} the polarization dependent part of the coupling as obtained from the solution of the full set equations of motions \eqref{eq:EOMfull}, for a small initial tilt of the collective nuclear spin in the linear response regime in the two configurations (solid lines), and from the solution of the simplified models \eqref{eq:conf1AdiabEl} in ``Config.1" (circles) and \eqref{eq:conf2AdiabEl} in ``Config.2" (squares). On the same plot, we show the analytic expressions of the functions \eqref{eq:f12} and \eqref{eq:f32}. Overall, the results in ``Config.2" show good agreement, while the results in ``Config.1" have a larger discrepancy especially for large polarizations. This is expected, as the effects of the $F=3/2$ manifold have been completely neglected.

Even accounting for the difference in the coupling constants in the two configurations, $\eta$ being approximately $0.48$ times $\chi$, due to the large difference in the the scaling factors  $f^{(3/2)}$ and $f^{(1/2)}$ the effective coupling between nuclear spin and light is significantly larger in ``Config.2" than in ``Config.1".

\section{Conclusions}
In conclusion, we have derived the full set of semiclassical equation of motion describing the interaction between light and metastable helium-3, taking into account metastability-exchange collisions with helium-3 atoms in the ground state as well as a static external magnetic field. We then explored interesting choices of detunings between light and metastable helium-3 $2^3S-2^3P$ transition, and found two configurations that are dominated by the interaction with the atomic $F=1/2$ or $F=3/2$ manifolds. At these configurations we were able to write linearised equations of motion that describe an effective Faraday interaction between light and helium-3 nuclear spin. We provide an expression for the effective coupling rate as a function of the experimental parameters, and conclude that for large nuclear spin polarization this quantity is considerably larger in the configuration dominated by the $F=3/2$ manifold. A comparison between the numerical solution of the full set of equations of motion and the linearised model show that for large spin polarizations the light field evolution contains contributions from the coupling with tensor spin components. In the future, it will be important to explore how the presence of these tensor contributions might affect squeezing of the nuclear spin in a fully quantum treatment \cite{AlanLong,AlanPRL}.

\vspace{10mm}
\textbf{Acknowledgments:} MF was supported by the Swiss National Science Foundation Ambizione Grant No. 208886, and The Branco Weiss Fellowship -- Society in Science, administered by the ETH Z\"{u}rich.

\clearpage
\newpage

\appendix
\section{Transition frequencies between the metastable and the excited states}
\label{app:table_fr}

Frequencies of transitions $C_1-C_9$ shown in Fig.\ref{fig:levels}, and calculated with respect to transition $C_8=2\pi \, \unit{276\, 726\, 257}{MHz}$ are reported in the following table \cite{Courtade}
% Table \ref{tab:Cfreqs}.

\begin{table}[h!]
    \centering
    \begin{tabular}{c|c|c|c|c|c}
               &    Freq. offset (GHz)  & $F$ & $J$ & $F^\prime$ & $J^\prime$ \\ 
        \hline
         $\Delta_{1}/2\pi$ &  -32.6045 & $1/2$ & $1$ & $3/2$ & $1$ \\
         $\Delta_{2}/2\pi$ &  -28.0929 & $1/2$ & $1$ & $1/2$ & $1$ \\
         $\Delta_{3}/2\pi$ &  -27.6453 & $3/2$ & $1$ & $5/2$ & $2$ \\
         $\Delta_{4}/2\pi$ &  -27.4238 & $1/2$ & $1$ & $3/2$ & $2$ \\
         $\Delta_{5}/2\pi$ &  -25.8648 & $3/2$ & $1$ & $3/2$ & $1$ \\
         $\Delta_{6}/2\pi$ &  -21.3532 & $3/2$ & $1$ & $1/2$ & $1$ \\
         $\Delta_{7}/2\pi$ &  -20.6841 & $3/2$ & $1$ & $3/2$ & $2$ \\
         $\Delta_{8}/2\pi$ &  0 & $1/2$ & $1$ & $1/2$ & $0$ \\
         $\Delta_{9}/2\pi$ & +6.7397 & $3/2$ & $1$ & $1/2$ & $0$
    \end{tabular}
    %\caption{}
    \label{tab:Cfreqs}
\end{table}

\section{Effective Hamiltonian in a state $F$ in the large detuning limit}
\label{app:Heff}
The Hamiltonian for a single-atom interacting with a light field is
\begin{align}\label{eq:HAL}
    h_F = & \sum_{F'} \hbar \dfrac{\sigma_{F'}}{2A} \dfrac{\Gamma/2}{\Delta_{F'}+i\Gamma/2} \left\{ \alpha_{F'}^V F_z S_z +  \right. \\ 
    & \left. + \dfrac{\alpha_{F'}^T}{(F+1)}\left[ \left( \dfrac{F(F+1)}{3} - F_z^2 \right) S_0 + (F_x^2-F_y^2)S_x + (F_x F_y + F_y F_x)S_y \right] \right\} \;.\notag
\end{align} 
In Eq.~\ref{eq:HAL}, $F$ ($F'$) is the total angular momentum of the starting (target) state of the transition, $\sigma_{F'}$ is the resonant effective cross section of the transition $F\rightarrow F'$, and $\Delta_{F'}=\omega_{\text{probe}}-\omega_{F F'}$ is the detuning with respect to the resonance. %($\Delta_{F'}>0$ if the light is detuned towards the blue of the transition).  
In the expression $\Delta_{F'}+i\Gamma/2$ in the denominator, the imaginary part can be neglected for $\Delta_{F'}\gg \Gamma/2$.

The vector and tensor components of the polarisation have the form \cite{Pinard07,GeremiaPRA06} \begin{align}
    \alpha_{F'}^V &= \dfrac{3(2J'+1)}{2(2F'+1)(2J+1)}\left(-\dfrac{2F-1}{F}\delta_{F-1}^{F'} - \dfrac{2F+1}{F(F+1)}\delta_{F}^{F'} + \dfrac{2F+3}{F+1}\delta_{F+1}^{F'} \right) \\
    \alpha_{F'}^T &= - \dfrac{3(F+1)(2J'+1)}{2(2F'+1)(2J+1)}\left(\dfrac{1}{F}\delta_{F-1}^{F'} - \dfrac{2F+1}{F(F+1)}\delta_{F}^{F'} + \dfrac{1}{F+1}\delta_{F+1}^{F'} \right) \,.
\end{align} Introducing $\sigma_2=3\lambda^2/2\pi$ and Wigner's $6j$ symbols $\{ \}$, the resonant effective cross section $\sigma_{F'}$ between two levels $F, J, I$ and $F', J', I$ is given by \begin{equation}
    \sigma_{F'} = \sigma_2 \dfrac{2(2J+1)(2F'+1)}{3} \begin{Bmatrix}
        J' & 1 & J\\
        F & I & F'
    \end{Bmatrix}^2 \;.
\end{equation} For a spin greater than $F=1/2$, the irreducible tensor basis $t^{l}_{m}$ should be used, with $l=0,1,..,2F$ and $m=-l,...,l$ defined as a function of the ladder operators $F_{\pm}=F_x \pm i F_y$, and given below for $l\le 3$.
\begin{subequations}\label{eq:Tdef}
\begin{align}
    t^0_0 &= n^0_0 \; \mathbb{I}\\  
    t^1_0 &= n^1_0 \; F_z \\  
    t^1_{\pm 1} &= n^1_{\pm 1} \; F_{\pm} \\
    t^2_0 &= n^2_0 \; (3 F_z^2 - \mathbf{F}^2 ) \\  
    t^2_{\pm 1} &= n^2_{\pm 1} \; (F_{\pm} F_z + F_z F_{\pm})\\  
    t^2_{\pm 2} &= n^2_{\pm 2} \; F_{\pm}^2 \\  
    t^3_0 &= n^3_0 \; (5 F_z^2 - 3 \mathbf{F}^2 + 1) F_z\\  
    t^3_{\pm 1} &= n^3_{\pm 1} \; \left[ (5 F_z^2 - \mathbf{F}^2 - 1/2) F_{\pm} + F_{\pm} (5 F_z^2 - \mathbf{F}^2 - 1/2) \right] \\ 
    t^3_{\pm 2} &= n^3_{\pm 2} \; (F_{\pm}^2 F_z + F_{\pm} F_z F_{\pm} + F_z F_{\pm}^2 ) \\ 
    t^3_{\pm 3} &= n^3_{\pm 3} \; F_{\pm}^3 
\end{align}
\end{subequations} 
As expected, for a spin $F$ operators of rank $l>2F$ are null. The operators $t^{l}_{m}$ satisfy property $(t^l_m)^\dagger = (-1)^m t^l_{-m}$ and are of null trace except $t^{l=0}_0$. Other properties and commutation relations are given in Appendix \ref{app:commF}. Prefactors $n^l_m$ are chosen to ensure the normalisation condition $\text{Tr}[t^l_m (t^l_m)^\dagger]=1$, and for a spin $F=3/2$ they read
\begin{table}[h!]
    \centering
    \begin{tabular}{c|c|c|c|c|c|c|c|c|c}
        $n^0_0$ & $n^1_0$ & $n^1_{\pm 1}$ & $n^2_0$ & $n^2_{\pm 1}$ & $n^2_{\pm 2}$ & $n^3_0$ & $n^3_{\pm 1}$ & $n^3_{\pm 2}$ & $n^3_{\pm 3}$  \\ \hline
        $\frac{1}{2}$ & $\frac{1}{\sqrt{5}}$ & $\mp\frac{1}{\sqrt{10}}$ & $\frac{1}{6}$ & $\mp\frac{1}{2\sqrt{6}}$ & $\frac{1}{2\sqrt{6}}$ & $\frac{1}{3\sqrt{5}}$ & $\mp\frac{1}{4\sqrt{15}}$ & $\frac{1}{3\sqrt{6}}$ & $\mp\frac{1}{6}$ 
    \end{tabular}
\end{table}
 
Finally, we introduce the symmetric and antisymmetric combinations \begin{equation}
    \mathfrak{R}t^{l}_{m} \equiv \dfrac{t^{l}_{m} + (t^{l}_{m})^\dagger }{\sqrt{2}} \qquad\qquad \mathfrak{I}t^{l}_{m} \equiv \dfrac{t^{l}_{m} - (t^{l}_{m})^\dagger }{i\sqrt{2}} \;\;.
\end{equation} 

The associated collective operators are then defined by summing over all particles as \eg $\mathfrak{R}T^{l}_{m}=\sum_{i=1}^N (\mathfrak{R}t^{l}_{m})_i$.
We note that $\TT{1}{0}$ is proportional to the longitudinal magnetisation, $\reToo$ and $\imToo$ to the magnetisations according to $x$ and $y$, $\TT{2}{0}$ is the quadrupolar spin polarisation (called alignment), $\reTto$ and $\imTto$ imply coherences between levels $\Delta m =1$, $\reTtt$ and $\imTtt$ between levels $\Delta m =2$, $\TT{3}{0}$ is the octopolar spin polarisation, etc. 

Assuming $\Delta \gg \Gamma$, and summing over all atoms, we can rewrite the collective Hamiltonian
\begin{equation}\label{eq:theHam}
    H_F = \sum_{F'} \hbar \dfrac{\sigma_{F'}}{4A} \dfrac{\Gamma}{\Delta_{F'}} \left\{ \alpha_{F'}^V F_z S_z + \dfrac{\alpha_{F'}^T}{(F+1)}\left[ - 2 \TT{2}{0} S_0 + \sqrt{12} \left( \reTtt S_x + \imTtt S_y \right) \right] \right\} \;,
\end{equation} where we have preferred the notation $F_z$ rather than $\TT{1}{0}$.

\section{Commutation relations for the atomic operators}
\label{app:commF}

Let $\vec{F}=\{F_x,F_y,F_z\}$ be the components of a spin operator, $F_{\pm}=F_x \pm i F_y$ the corresponding ladder operators, and $t^l_m$ the irreducible tensors explicit in (\ref{eq:Tdef}) for $l \le 3$. The operators satisfy the following commutator rules \cite{Pinard07,Siminovitch}~: \begin{align}
    [F_x, F_y] &= i F_z \qquad\text{(and cyclic permutations)} \\
    [F_z, F_{\pm}] &= \pm F_{\pm} \\
    [F_+, F_-] &= 2 F_z \\
    [F_z, t^{l}_{m}] &= m t^{l}_{m} \\
    [F_{\pm}, t^{l}_{m}] &= \sqrt{(l\pm m + 1)(l \mp m)} t^l_{m\pm 1}
\end{align} \begin{equation}
    [ t^{l_1}_{m_1}, t^{l_2}_{m_2} ] = \sum_{L,M} (-1)^{L+2F} \sqrt{(2l_1+1)(2l_2+1)} \begin{Bmatrix}
        l_1 & l_2 & L\\
        F & F & F
    \end{Bmatrix} \langle l_1 m_1 l_2 m_2 , L M \rangle [1-(-1)^{l_1+l_2+L}] \; t^L_M \;,
\end{equation} where $\{\}$ denotes Wigner's 6j symbols, and $\langle,\rangle$ the Clebsch-Gordan coefficients.

\section{Derivation of metastability exchange equations}
\label{app:deivMEC}

In this appendix we explain how the metastability exchange equations presented in section \ref{sub:eomMEC} can be derived in practice.
First, the density matrix $\rho$ can be written as $\rho=\sum_{i,j}\rho_{i,j}\ket{i}\bra{j}$, where the indices $i,j$ label the basis states 
\begin{subequations}\label{eq:MixedSt}
\begin{align}
    \ket{5} = \sqrt{\dfrac{1}{3}} \ket{0,-\dfrac{1}{2}} - \sqrt{\dfrac{2}{3}} \ket{-1,\dfrac{1}{2}} &\; \ket{6} = -\sqrt{\dfrac{1}{3}} \ket{0,\dfrac{1}{2}} + \sqrt{\dfrac{2}{3}} \ket{1,-\dfrac{1}{2}} \\
        \ket{1} =\ket{-1,-\dfrac{1}{2}} \; \ket{2} = \sqrt{\dfrac{2}{3}} \ket{0,-\dfrac{1}{2}} + \sqrt{\dfrac{1}{3}} \ket{-1,\dfrac{1}{2}} &\; \ket{3} = \sqrt{\dfrac{2}{3}} \ket{0,\dfrac{1}{2}} + \sqrt{\dfrac{1}{3}} \ket{1,-\dfrac{1}{2}} \; \ket{4} =\ket{1,\dfrac{1}{2}} \\
        \ket{9} =\ket{-\dfrac{1}{2}} &\; \ket{0} =\ket{\dfrac{1}{2}} \;.
\end{align} 
\end{subequations}
Here, note that $\ket{9}$ and $\ket{0}$ are purely nuclear states, while the others are hyperfine states of total spin $F=3/2$ and $F=1/2$ expressed in the decoupled basis of the electronic and nuclear spin. In practice, we neglect coherences between metastable and ground states, as well as coherences between the $3/2$ and $1/2$ states. This gives us
\begin{equation}
\rho = \left[\begin{array}{ c | c}
    \rho_m & 0\\
    \hline
    0 & \rho_f
  \end{array}\right] =
\left[
\begin{array}{cccccc | cc}
 \rho_{1,1} & \rho_{1,2} & \rho_{1,3} & \rho_{1,4} & 0 & 0 & 0 & 0 \\
 \rho_{2,1} & \rho_{2,2} & \rho_{2,3} & \rho_{2,4} & 0 & 0 & 0 & 0\\
 \rho_{3,1} & \rho_{3,2} & \rho_{3,3} & \rho_{3,4} & 0 & 0 & 0 & 0\\
 \rho_{4,1} & \rho_{4,2} & \rho_{4,3} & \rho_{4,4} & 0 & 0 & 0 & 0\\
 0 & 0 & 0 & 0 & \rho _{5,5} & \rho _{5,6} & 0 & 0\\
 0 & 0 & 0 & 0 & \rho _{6,5} & \rho _{6,6} & 0 & 0\\
\hline
    0 & 0 & 0 & 0 & 0 & 0 & \rho _{9,9} & \rho _{9,0}  \\
    0 & 0 & 0 & 0 & 0 & 0 & \rho _{0,9} & \rho _{0,0}
  \end{array}\right] \;.
\end{equation}
Using now Eqs.~(\ref{eq:rhoMEC1},\ref{eq:rhoMEC2}), with Eqs.~(\ref{eq:rhoPT1},\ref{eq:rhoPT2}), allows us to derive the equations of motion of $\rho$ due to metastability exchange collisions. These can be found in the appendix of Ref.~\cite{Reinaudi}. Finally, the evolution due to metastability exchange collisions of any one-body atomic operator $O$ can be calculated using Eq.~\eqref{eq:eomOpMEC}. To this end, it is convenient to express the spin operators in the basis Eq.~\eqref{eq:MixedSt}. For the $F=1/2$ metastable state and the nuclear ground state, the spin operators in the relevant $2\times 2$ subspace are proportional to the Pauli matrices, \eg $k_z = \frac{1}{2}\left(\begin{smallmatrix}
    -1 & 0\\
    0 & 1
\end{smallmatrix}\right)$. For the $F=3/2$ metastable state the spin operators in the relevant $4\times 4$ subspace are
\begin{align}
    j_x = \left(
\begin{array}{cccc}
 0 & \frac{\sqrt{3}}{2} & 0 & 0 \\
 \frac{\sqrt{3}}{2} & 0 & 1 & 0 \\
 0 & 1 & 0 & \frac{\sqrt{3}}{2} \\
 0 & 0 & \frac{\sqrt{3}}{2} & 0 \\
\end{array}
\right) \;, \qquad
    j_y = \left(
\begin{array}{cccc}
 0 & \frac{i \sqrt{3}}{2} & 0 & 0 \\
 -\frac{i \sqrt{3}}{2} & 0 & i & 0 \\
 0 & -i & 0 & \frac{i \sqrt{3}}{2} \\
 0 & 0 & -\frac{i \sqrt{3}}{2} & 0 \\
\end{array}
\right) \;, \qquad
    j_z = \left(
\begin{array}{cccc}
 -\frac{3}{2} & 0 & 0 & 0 \\
 0 & -\frac{1}{2} & 0 & 0 \\
 0 & 0 & \frac{1}{2} & 0 \\
 0 & 0 & 0 & \frac{3}{2} \\
\end{array}
\right) \;, 
\end{align}
\begin{align}
t^{2}_{0} = \left(
\begin{array}{cccc}
 \frac{1}{2} & 0 & 0 & 0 \\
 0 & -\frac{1}{2} & 0 & 0 \\
 0 & 0 & -\frac{1}{2} & 0 \\
 0 & 0 & 0 & \frac{1}{2} \\
\end{array}
\right) \;,\qquad
\mathfrak{R}t^{2}_{1} = \left(
\begin{array}{cccc}
 0 & \frac{1}{2} & 0 & 0 \\
 \frac{1}{2} & 0 & 0 & 0 \\
 0 & 0 & 0 & -\frac{1}{2} \\
 0 & 0 & -\frac{1}{2} & 0 \\
\end{array}
\right) \;,\qquad
\mathfrak{R}t^{2}_{2} = \left(
\begin{array}{cccc}
 0 & 0 & \frac{1}{2} & 0 \\
 0 & 0 & 0 & \frac{1}{2} \\
 \frac{1}{2} & 0 & 0 & 0 \\
 0 & \frac{1}{2} & 0 & 0 \\
\end{array}
\right) \;,\quad
\end{align}
\begin{align}
\mathfrak{I}t^{2}_{1} = \left(
\begin{array}{cccc}
 0 & \frac{i}{2} & 0 & 0 \\
 -\frac{i}{2} & 0 & 0 & 0 \\
 0 & 0 & 0 & -\frac{i}{2} \\
 0 & 0 & \frac{i}{2} & 0 \\
\end{array}
\right) \;, \quad
\mathfrak{I}t^{2}_{2} = \left(
\begin{array}{cccc}
 0 & 0 & \frac{i}{2} & 0 \\
 0 & 0 & 0 & \frac{i}{2} \\
 -\frac{i}{2} & 0 & 0 & 0 \\
 0 & -\frac{i}{2} & 0 & 0 \\
\end{array}
\right) \;,\quad
    t^{3}_{0} = \left(
\begin{array}{cccc}
 -\frac{1}{2 \sqrt{5}} & 0 & 0 & 0 \\
 0 & \frac{3}{2 \sqrt{5}} & 0 & 0 \\
 0 & 0 & -\frac{3}{2 \sqrt{5}} & 0 \\
 0 & 0 & 0 & \frac{1}{2 \sqrt{5}} \\
\end{array}
\right) \;,\quad
\end{align}
\begin{align}
\mathfrak{R}t^{3}_{1} = \left(
\begin{array}{cccc}
 0 & -\frac{1}{\sqrt{10}} & 0 & 0 \\
 -\frac{1}{\sqrt{10}} & 0 & \sqrt{\frac{3}{10}} & 0 \\
 0 & \sqrt{\frac{3}{10}} & 0 & -\frac{1}{\sqrt{10}} \\
 0 & 0 & -\frac{1}{\sqrt{10}} & 0 \\
\end{array}
\right) \;,\quad
\mathfrak{R}t^{3}_{2} = \left(
\begin{array}{cccc}
 0 & 0 & -\frac{1}{2} & 0 \\
 0 & 0 & 0 & \frac{1}{2} \\
 -\frac{1}{2} & 0 & 0 & 0 \\
 0 & \frac{1}{2} & 0 & 0 \\
\end{array}
\right) \;,\quad
\mathfrak{R}t^{3}_{3} = \left(
\begin{array}{cccc}
 0 & 0 & 0 & -\frac{1}{\sqrt{2}} \\
 0 & 0 & 0 & 0 \\
 0 & 0 & 0 & 0 \\
 -\frac{1}{\sqrt{2}} & 0 & 0 & 0 \\
\end{array}
\right) \;,\quad
\end{align}
\begin{align}
\mathfrak{I}t^{3}_{1} = \left(
\begin{array}{cccc}
 0 & -\frac{i}{\sqrt{10}} & 0 & 0 \\
 \frac{i}{\sqrt{10}} & 0 & i \sqrt{\frac{3}{10}} & 0 \\
 0 & -i \sqrt{\frac{3}{10}} & 0 & -\frac{i}{\sqrt{10}} \\
 0 & 0 & \frac{i}{\sqrt{10}} & 0 \\
\end{array}
\right) \;,\quad
\mathfrak{I}t^{3}_{2} = \left(
\begin{array}{cccc}
 0 & 0 & -\frac{i}{2} & 0 \\
 0 & 0 & 0 & \frac{i}{2} \\
 \frac{i}{2} & 0 & 0 & 0 \\
 0 & -\frac{i}{2} & 0 & 0 \\
\end{array}
\right) \;,\quad
\mathfrak{I}t^{3}_{3} = \left(
\begin{array}{cccc}
 0 & 0 & 0 & -\frac{i}{\sqrt{2}} \\
 0 & 0 & 0 & 0 \\
 0 & 0 & 0 & 0 \\
 \frac{i}{\sqrt{2}} & 0 & 0 & 0 \\
\end{array}
\right) \;.
\end{align}

The equations of motion for the collective spin operators are then readily found by taking
\begin{align}
    \avg{\vec{I}} = N \avg{\vec{i}} \;,\qquad \avg{\vec{K}} = n \avg{\vec{k}} \;,\qquad \avg{\vec{J}} = n \avg{\vec{j}} \;,\qquad \avg{\reTtt} = n \avg{\mathfrak{R}t^{2}_{2}} \;,\qquad \text{etc.}
\end{align}
and are the one presented in Section~\ref{sub:eomMEC}.

\section{Coefficients of the linearized equations of the simplified models}
\label{app:coeffsAdiabEl}

In this appendix we give the expression of the coefficients appearing in the linearized equations of the simplified models of section \ref{sec:configs}.

The coefficients appearing in Eqs.~\eqref{eq:conf1AdiabEl} read
\begin{align}
    c^{(1/2)} &= \left(\frac{30 i \left(M^2+4\right) \gamma_{\text{3/2}} B_x}{\gamma _m}+\frac{27 \left(M^2+4\right)^2 \gamma _{\text{3/2} \left(M^2-8\right)}^2
   B_x^2}{\gamma _m^2}+\left(M^2-8\right)^2\right) \\
   a_1^{(1/2)} &= \frac{\left(M^2-8\right)}{(M^2+4)(M^2-1)} \left(\left(M^2+3\right) c^{(1/2)} + 4 \left(\left(M^2-8\right) \left(2 M^2+5\right)-\frac{3 i
   \left(M^2+4\right) \left(M^2+5\right) \gamma _{\text{3/2}} B_x}{2 \gamma
   _m}\right)\right) \\
   a_2^{(1/2)} &= \left(\frac{9 \left(M^2+4\right) \gamma _{\text{3/2}}^2
   B_x^2}{\gamma _m^2}+\left(M^2-8\right)^2\right) \\
   b_1^{(1/2)} &= \left(-\frac{21 \left(M^2+4\right) \gamma
   _{\text{3/2}}^2 B_x^2}{\gamma _m^2}-\frac{2 i \left(M^4+2 M^2-80\right) \gamma
   _{\text{3/2}} B_x}{\gamma _m}+\left(M^2-8\right)^2\right) \\
   b_2^{(1/2)} &= \left(\frac{9 \left(M^2+4\right) \gamma
   _{\text{3/2}}^2 B_x^2}{\gamma _m^2}+\left(M^2-8\right)^2\right)
\end{align}

The coefficients appearing in Eqs.~\eqref{eq:conf2AdiabEl} read
\begin{align}
    c^{(3/2)} &= \frac{6 i \left(M^2+4\right) B_x \left(6 \gamma _{\text{1/2}}+7 \gamma
   _{\text{3/2}}\right)}{\gamma _m}-\frac{27 \left(M^2+4\right)^2 \gamma _{\text{1/2}}
   \gamma _{\text{3/2}} B_x^2}{\left(M^2+7\right) \gamma _m^2}+8 \left(M^2+7\right)^2 \\
   a_1^{(3/2)} &= \frac{\left(M^2+7\right)}{\left(M^2+4\right) \left(M^2+5\right)} \left(\left(M^2+3\right) c^{(3/2)} - 8 \left( \left(M^2+1\right)
   \left(M^2+7\right)-\frac{3 i \left(M^2-1\right) \left(M^2+4\right) \gamma _{\text{3/2}}
   B_x}{4 \gamma _m}\right)\right) \\
   a_2^{(3/2)} &= 2 \left(\frac{12 i \left(M^2+7\right) B_x \left(\gamma _{\text{1/2}}+\gamma
   _{\text{3/2}}\right)}{\gamma _m}-\frac{9 \left(M^2+4\right) \gamma _{\text{1/2}} \gamma
   _{\text{3/2}} B_x^2}{\gamma _m^2}+4 \left(M^2+7\right)^2\right) \\
   a_3^{(3/2)} &= \frac{24 M^2 \left(M^2+7\right)}{(M^2+5)} \\   
   b_1^{(3/2)} &= -4 \left(\frac{i \left(M^2+7\right) B_x \left(\left(M^2-8\right) \gamma _{\text{1/2}}-6
   \gamma _{\text{3/2}}\right)}{\gamma _m}+\frac{6 \left(M^2+4\right) \gamma _{\text{1/2}}
   \gamma _{\text{3/2}} B_x^2}{\gamma _m^2}-2 \left(M^2+7\right)^2\right) \\
   b_2^{(3/2)} &= a_2^{(3/2)} + a_3^{(3/2)} \frac{i B_x \left(\gamma _{\text{1/2}}+\gamma
   _{\text{3/2}}\right)}{\gamma _m}\\
   b_3^{(3/2)} &= \frac{12 M^2}{(M^2+5)} \left(-2 \left(M^2+7\right)+\frac{3 i \left(M^2+4\right) \gamma _{\text{1/2}}
   B_x}{\gamma _m}\right)
\end{align}

\clearpage
\newpage
\section{Adiabatic elimination in Configuration 2}
\label{app:adiabNum}

In Section~\ref{sec:Conf2} we obtain for Configuration 2 a set of simple equations of motion for the variables $S$, $I$ and $J$ by first setting the coupling coefficients $\chi=\mu=0$ in Eqs.~\eqref{eq:linEOMfull}, and then adiabatically eliminating the $\delta K_\alpha$ and $\delta Q_{\alpha x}$ degrees of freedom.

\begin{figure}
    \centering
    \includegraphics[width=\textwidth]{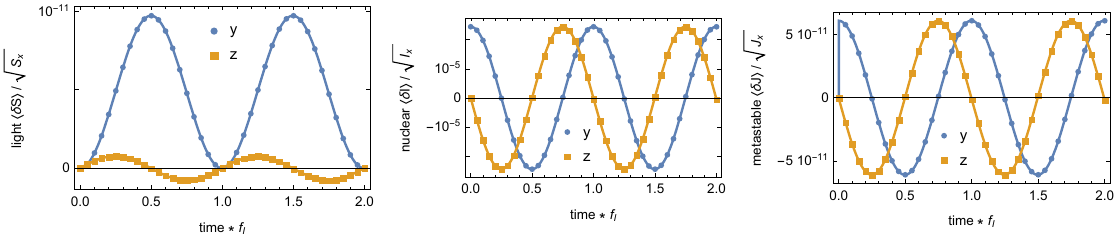}
    \caption{Comparison between the full model Eq.~\eqref{eq:EOMfull} and the same model after adiabatic elimination of the $F=1/2$ and tensor degrees of freedom in ``Config.2". The nuclear magnetization is set to $M=98\%$. Blue indicates the $y$ spin component, while yellow the $z$ spin component. Solid lines are the full model, while round and square markers indicate the numerical solution of the same model after adiabatic elimination of the $F=1/2$ and tensor degrees of freedom (without setting $\mu=\chi=0$, as done in Fig.~\eqref{fig:CompConf2}).
    Simulation parameters: $n_\text{ph}/N_{\rm cell}=10^{-3}$, $N_{\text{cell}}/n_{\text{cell}} = T/\tau=\unit{10^{6}}{s}$, and an initial state resulting from tilting the nuclear spin by $\unit{0.01}{rad}$. Time is in units of the nuclear spin Larmor frequency.
    %Simulation parameters: $B_x=\unit{5}{mG}$, $n_\text{ph}=10^3$, $T=\unit{1}{s}$, $\tau=\unit{10^{-6}}{s}$, $N=10^6$, $n=1$, and an initial state resulting from tilting the nuclear spin by $\unit{0.01}{rad}$.
    }
    \label{fig:AdiabNoApprox}
\end{figure}

Here, we perform instead the adiabatic elimination of $\delta K_\alpha$ and $\delta Q_{\alpha x}$ directly on Eqs.~\eqref{eq:linEOMfull}, keeping all the coupling coefficients. This lead to a set of expressions that are too complex to be treated analytically, but can be solved numerically showing good agreement with the full model, see Fig.~\ref{fig:AdiabNoApprox}.

\begin{thebibliography}{}
\bibitem{footnoteStokes} Note that in \cite{Pinard07} there is an additional 1/2 factor, that in our case has been included in the definition of the Stokes operators.
\end{thebibliography}
\end{document}